\documentclass[11pt,letter,reqno]{amsart}
\usepackage{amsthm, amsmath, amssymb, amsfonts, graphicx, epsfig}
\usepackage{accents}

\usepackage{bbm}
\usepackage{mathrsfs,dsfont}

\usepackage[normalem]{ulem}
\usepackage{cancel}

\usepackage{graphicx}
\usepackage[font=sl,labelfont=bf]{caption}
\usepackage{subcaption}

\usepackage{float}
\restylefloat{table}

\usepackage{enumerate}
\usepackage[colorlinks=true, pdfstartview=FitV, linkcolor=blue,
            citecolor=blue, urlcolor=blue]{hyperref}
\usepackage[usenames]{color} 

\usepackage[round]{natbib}

\usepackage{url}
\makeatletter\def\url@leostyle{%
 \@ifundefined{selectfont}{\def\UrlFont{\sf}}{\def\UrlFont{\scriptsize\ttfamily}}} \makeatother

\newtheorem{theorem}{Theorem}

\theoremstyle{definition}

\newtheorem{example}[theorem]{Example}
\theoremstyle{remark}
\newtheorem{remark}[theorem]{Remark}

\numberwithin{equation}{section}
\numberwithin{theorem}{section}

\definecolor{Red}{rgb}{0.8,0,0.1}

\def\cA{\mathcal{A}}

\def\cN{\mathcal{N}}

\def\bE{E}
\def\bF{\mathbb{F}}

\def\bN{\mathbb{N}}

\def\bR{\mathbb{R}}

\newcommand{\1}{\mathbbm{1}}            

\DeclareMathOperator*{\argmin}{arg\,min} 

\DeclareMathOperator{\var}{\mathrm{V}@\mathrm{R}}           
\DeclareMathOperator{\evar}{\mathrm{EV}@\mathrm{R}}           

\makeatletter
\def\namedlabel#1#2{\begingroup
    #2%
    \def\@currentlabel{#2}%
    \phantomsection\label{#1}\endgroup
}
\makeatother

\usepackage{multirow}

\title[Estimating and backtesting risk]{Estimating and backtesting  risk under heavy tails}

\DeclareMathOperator{\ES}{ES}

\usepackage[backgroundcolor=white,bordercolor=orange]{todonotes}
\usepackage{booktabs,nicefrac}
\usepackage{array}
\usepackage{tcolorbox}

\author{
        Marcin Pitera \and Thorsten Schmidt}
 \address{Institute of Mathematics, Jagiellonian University, \L{}ojasiewicza 6, 30-348 Cracow, Poland} \email{\url{marcin.pitera@im.uj.edu.pl}}
 \address{Dep.~of Mathematical Stochastics, University of Freiburg, Eckerstr.1, 79104 Freiburg, Germany}
   \email{\url{thorsten.schmidt@stochastik.uni-freiburg.de}}

\binoppenalty=\maxdimen 
\relpenalty=\maxdimen

\date{First circulated: October 20, 2020, This version: \today}

\newcommand{\R}{\bR}

\defcitealias{Bas2013}{BCBS, 2013} 
\defcitealias{Bas2011}{BCBS, 2011}
\defcitealias{Bas2006}{BCBS, 2006} 
\defcitealias{Bas1996}{BCBS, 1996}

\usepackage{caption}
\begin{document}

\maketitle

\begin{abstract}

While the {estimation} of risk is an important question in the daily business of banking and insurance, many existing plug-in estimation procedures suffer from an unnecessary bias. This often leads to the underestimation of risk and negatively impacts backtesting results, especially in small sample cases. In this article we show that the link between estimation bias and backtesting  can be traced back to the dual relationship between risk measures and the corresponding performance measures, and discuss this in reference to value-at-risk, expected shortfall and expectile value-at-risk.

Motivated by the consistent underestimation of risk by plug-in procedures, we propose a new algorithm for bias correction and show how to apply it for generalized Pareto distributions to the i.i.d.\ setting and to a GARCH(1,1) time series. In particular, we show that the application of our algorithm leads to gain in efficiency when heavy tails or heteroscedasticity exists in the data.
\\
\end{abstract}


\section{Introduction}

Risk measures are a central tool in capital reserve evaluation and quantitative risk management, see \cite{MFE} and references therein.  Since efficient statistical estimation procedures for risk are highly important in practice, this aspect of risk measurement recently raised a lot of attention.  Already \cite{ConDegSca2010} observed that sensitivities of risk estimations with respect to the underlying data set pose a challenging problem when robustness is considered. \cite{KraSchZah2014} showed that a certain notion of statistical robustness for law-invariant convex risk measures always holds. The backtesting aspect and its relation to elicitability has been discussed  in 
\cite{Acerbi2014Risk}, \cite{Davis2016},  and  \cite{Ziegel2014}, for example. 
Risk estimation statistical properties are also discussed in \cite{BarTan2021} and \cite{LauZah2016,LauZah2017}. The latter work also considers the \emph{statistical bias} of estimators, not to be confused with the risk bias we consider below in Equation \eqref{eq:rho.unbiased} following \cite{PitSch2016}, see also \cite{Frank2016} and \cite{FraHer2012} for the necessity to account for the risk bias.
Moreover, \cite{YueStoCoo2020} considered extreme value-at-risk in the context of distributionally robust inference.

When the estimation of risk is considered, the  majority of the literature focuses on the {\it plug-in} approach combined with  standard estimation techniques like maximum-likelihood estimation, see \cite{embrechts2014}, \cite{KraZah2017} and references therein. In this  context, topics like predictive statistical inference analysis and bias quantification are typically studied in reference to the fit of the whole estimated distribution. Still, for risk measurement purposes, it is also important to measure adequacy and conservativeness of the capital projections directly, e.g.\ via backtesting. We refer to  \cite{NolZie2017} where the importance of such studies is shown in reference to elicitability and backtesting. The aim of this article is to consider this topic in reference to common distributional assumptions including the ones with heavy tails.

In particular, the risk of underestimating the required regulatory capital, which is embedded into the risk estimation process, and  theoretical links between risk estimation  and backtesting performance should be further studied; see e.g.~\cite{BigTsa2016} or \cite{GerTsa2011}. To justify this point, let us mention that it was only recently pointed out in \cite{MolPit2017} that the standard value-at-risk (VaR) backtesting breach count statistic is in fact a performance measure dual to the VaR family of risk measures as introduced in \cite{CheMad2009}. This shows that the VaR breach count statistic is bound to the VaR family in the same manner like the Sharpe ratio is bound to risk aversion specification in mean-variance portfolio optimization or acceptability indices are bound to coherent risk measures, see \cite{BieCiaDraKar2013}. This indicates that there should be a direct statistical connection between the way how risk measure estimators  are designed and how they perform in backtesting.

In this article, we study this connection by discussing estimation bias in the context of estimating  risk  and its connection to backtesting performance. We focus on an economic notion of unbiasedness for risk measures which turns out to be important whenever the underlying model needs to be estimated. Unbiasedness in the context of risk measures is the generalization of the well-known statistical unbiasedness to the risk landscape, see \cite{PitSch2016}.  Also, we refer to \cite{BigTsa2016} where a similar approach has been applied to residual risk quantification.

Here, we link the concept of unbiasedness for risk estimators  to backtesting and compare the performance of existing  estimators to the proposed  unbiased counterparts in Gaussian and generalized Pareto distribution (GPD) frameworks. The results presented for value-at-risk and expected shortfall show that plug-in estimators suffer from  a systematic underestimation of risk capital which results in deteriorated backtesting performance. While this is true even in an i.i.d.~ and light-tailed  setting, it is more pronounced for heavy tails, small sample sizes, in the presence of heteroscedasticity, and with high confidence levels. Also, we propose a new bias reduction technique that increases the efficiency of value-at-risk and expected shortfall plug-in estimators. 

This article is organized as follows. In Section \ref{sec:2} we introduce the theoretical background, while in Section~\ref{S:2} we comment on the relation between risk bias and backtesting. In Section \ref{ex:VAR.bt} and Section \ref{ex:ES.bt} we illustrate this relation for value-at-risk and expected shortfall. Next, in Section~\ref{S:new.one} we study backtesting in a more general context. In Section~\ref{S:bias.reduction}, we introduce a bootstrapping algorithm for bias reduction. Section \ref{sec:4} illustrates the performance on simulated examples under GPD distribution and shows that the risk bias correction leads to better estimator performance. Concluding remarks are provided in Section~\ref{S:conclusions}.

\section{The estimation of risk and the associated bias}\label{sec:2}

Consider the risk of a financial position  measured by a monetary risk measure $\rho$. The   position's future P\&L is denoted by $X$.  The estimation relies on a  sample $X_1,\dots,X_n$ corresponding to historic realizations of the P\&L.  An \emph{estimator} of the risk $\rho(X)$ is  a  measurable function of the  historical data, which we denote by $\hat \rho = \hat \rho(X_1,\dots,X_n)$. The position is  \emph{secured} by adding the estimated risk $\hat \rho$ to the position and we denote the  {\it secured position}  by
\begin{align}\label{eq:secured position}
	Y :=  X+\hat \rho.
\end{align}
Note that $Y$ is a random variable since both $X$ and $\hat \rho$ are random. 


Assuming the monetary risk measure $\rho$ is law-invariant, the risk of $Y$ depends on the underlying distribution of $(X_1,\ldots,X_n,X)$. Let $\Theta$ denote the parameter space and let each $\theta\in \Theta$ identify a certain distribution choice; $\Theta$ could be infinite-dimensional when linked to a non-parametric framework. For each $\theta\in\Theta$ we denote by $\rho_{\theta}(\cdot)$ the risk measure quantifying the underlying risk (e.g.~of~$Y$) under $\theta\in\Theta$. In particular,  $\rho(\cdot)=\rho_{\theta_0}(\cdot)$, where $\theta_0\in\Theta$ is the true but unknown parameter.


Following \cite{PitSch2016}, we call an estimator $\hat \rho$ \emph{unbiased} if 
\begin{equation}\label{eq:rho.unbiased}
\rho_{\theta}(X+\hat \rho)=0 \qquad \text{for all }\theta \in \Theta;
\end{equation}
if we want to emphasize the difference to statistical notion of unbiasedness, we use the term {\it risk unbiased} instead of {\it unbiased}.

The definition of unbiasedness in \eqref{eq:rho.unbiased} has a direct economic interpretation:  the secured position has to be acceptable under all  $\theta\in \Theta$. This quantifies the estimation error arising from  $\hat\rho$ and contains the information on the actual risk the holder of the position faces when using the estimator $\hat \rho$ to secure the future financial position $X$. Note this is aligned with the predictive inference paradigm as we quantify the actual risk of the secured future position when using estimated risk for securitization.

 We also refer to \cite{BigTsa2016}, where a concept of {\it residual risk} measuring the size of $\rho_{\theta}(X+\hat \rho)$ is introduced, and to \cite{FraHer2012} where so called {\it probability unbiasedness} is introduced -- those concepts are consistent with risk unbiasedness in the VaR case.

\begin{remark}[Relation to statistical bias]\label{rem:statistical.bias}
When the risk measure $\rho$ is linear, unbiasedness coincides with statistical bias (up to the sign): indeed, if $\rho_\theta = -E_\theta$, then \eqref{eq:rho.unbiased} is equivalent to 
$$E_\theta[\hat \rho]=-E_\theta[X], \quad \text{  for all }\theta\in\Theta, $$
where  $E_{\theta}$ is the expectation under the parameter $\theta\in\Theta$. 
In the i.i.d.~case,  the arithmetic mean multiplied by $-1$ turns out to be unbiased, just as in the statistical case. Also the estimator $\hat\rho:=-X_1$ is unbiased, but certainly not an optimal choice. This highlights why additional properties besides unbiasedness are important.
\end{remark}
 
The concept of unbiasedness we study here might be considered as a generalization of the statistical bias to non-linear measures. Moreover, it turns out to be important to require additional properties. For example,  consistency or other economic-driven properties such as minimization of the average mean capital reserve could be asked for. We refer to Example 7.3 in \cite{PitSch2016} for further details.

\begin{example}[Value-at-risk under normality]\label{ex:Gaussian}
Value-at-risk is the most recognised risk measure in both the financial and the insurance industry. For an arbitrary real-valued random variable $Z$ its value-at-risk at level $\alpha\in (0,1)$ is given by
\begin{equation}\label{eq:var}
\var_{\alpha}(Z):=\inf \{ x\in \bR\colon P(Z+x<0)\leq \alpha\}. 
\end{equation}
For the moment consider the simplest case, where  $X_1,\dots,X_n,X$ are i.i.d. and $X\sim \cN(\mu_0,\sigma_0^2)$. Then the true parameter $\theta_0 = (\mu_0,\sigma_0^2)$ is an element of the parameter space $\Theta = \R \times \R_{>0}$. The true value-at-risk at level $\alpha\in (0,1)$ of the position $X$ given $\theta_0$ is
\begin{align} \label{eq:varTrueNormal}
\var_{\alpha}(X)=-\left(\mu_0 + \sigma_0\,\Phi^{-1}(\alpha)\right), 
\end{align}
where $\Phi$ is the cumulative standard normal distribution function. We obtain that the value-at-risk of the secured position $Y$ vanishes, as it should be.

However, since $\theta_0$ is not known, it needs to be estimated and the picture changes. If we denote by $\hat \mu_n$ and $\hat \sigma_n$ the maximum-likelihood estimators based on the sample $X_1,\dots,X_n$ and plug them into Equation \eqref{eq:varTrueNormal}, we obtain the plug-in estimator
\begin{align} 
\hat\var^{\textrm{plug-in}}_{\alpha}=-\left(\hat \mu_n + \hat \sigma_n\,\Phi^{-1}(\alpha)\right).
\end{align}
In comparison to the estimation of a confidence interval in the Gaussian case, where a $t$-distribution arises, it is quite intuitive that the normal quantile should be replaced by a $t$-quantile to capture the uncertainty remaining in the parameter estimates $\hat \mu_n$ and $\hat \sigma_n$.
Indeed, it is not difficult to see that  the {\it unbiased estimator}  is given by
\begin{equation}\label{eq:unbiased.pre}
\hat \var^{\textrm{u}}_{\alpha}  :=-\Big(\hat\mu_n +\hat\sigma_n \sqrt{\frac{n+1}{n}}t_{n-1}^{-1}(\alpha)\Big),
\end{equation}
where $t_{n-1}$ refers to Student $t$-distribution function with $n-1$ degrees of freedom, see \cite{PitSch2016}.
Hence, unbiasedness requires that the risk of the  position $Y$ secured by the unbiased estimator $\hat \var^{\textrm{u}}_{\alpha}(X)$ vanishes for all $\theta \in \Theta$, as stated in Equation \eqref{eq:rho.unbiased}. We refer to Section~\ref{S:var1} for a discussion on how this impacts backtesting results.
\end{example}

As already mentioned, unbiasedness requires that the risk of the secured position is zero in \emph{all} scenarios, which turns out to be too demanding in some cases. Alternatively we consider also the case where the estimated risk capital is sufficient to insure the risks (but is possibly not the smallest capital choice). 
 
 In this regard, we call  $\hat \rho$  \emph{sufficient} if 
\begin{align}\label{eq:sufficient}
	\rho_{\theta}(X+\hat \rho)\le 0,\quad\textrm{for all $\theta \in \Theta$}.	
\end{align}
As previously, if we want to emphasize the difference to statistical notion of sufficiency, we will use the term {\it risk sufficient}.

\subsection{The law-invariant case}

The  family of risk measures  $(\rho_{\theta})_{\theta\in\Theta}$ is called \emph{law-invariant} if we find a function $R$ from the convex space of cumulative distribution functions (cdfs) to $\R\cup\{+\infty\}$, such that 
$$ \rho_\theta(Z) = R(F_Z(\theta))$$
for all $\theta \in \Theta$ and all $Z \in L^0$; here $F_Z(\theta) = P_\theta(Z \le \cdot)$ denotes the cdf of $Z$ under the probability measure associated to $\theta$. Value-at-risk and expected shortsall are prominent examples. In particular, we get $\rho_{\theta}(Y)=R(F_\theta(Y))$ for law-invariant families of risk measures.

Consequently,  the law-invariant estimator $\hat \rho$ is \emph{unbiased}  if 
\begin{equation}\label{eq:intro111}
R(F_{X+\hat\rho}(\theta))=0, \qquad \theta \in \Theta
\end{equation}
and sufficient if  $R(F_{X+\hat\rho}(\theta)) \le 0$ for all $\theta \in \Theta$.
In the i.i.d.~case $F_{X+\hat\rho}(\theta)$ can be obtained from the convolution of the distribution of $X$ with the distribution of the estimator $\hat \rho$. In the Gaussian case considered in Example \ref{ex:Gaussian}, this representation allowed us to quickly obtain the unbiased estimator in Equation \eqref{eq:unbiased.pre}.


%
%
%


\begin{remark}[Plug-in estimators]\label{rem:plugin} A common way to estimate risk measures is to use the plug-in procedure (see  Example \ref{ex:Gaussian}). In the case of a law-invariant risk measure this can be described as follows: first, estimate the unknown parameter by $\hat \theta$. Second, plug this estimator into the formula for $\rho_\theta(X)$, i.e.\ compute
$$ \hat \rho^{\textrm{plug-in}} = R(F_X(\hat \theta)). $$
Since the function $\theta \mapsto R(F_X(\theta))$ is typically highly non-linear, finite-sample properties of the estimators are often not inherited by $\hat \rho^{\textrm{plug-in}}$. This provides further motivation for studying unbiasedness of estimators of risk measures.
\end{remark}

%
%
%

\section{Relation between Backtesting and Bias}\label{S:2}
The notions of {\it unbiasedness} and {\it sufficiency} have an intrinsic connection to backtesting. In particular, the backtesting framework representation from \cite{MolPit2017} can be placed into the context of risk unbiasedness which we illustrate in the following.

\subsection{Backtesting}
Backtesting is a well-established procedure of checking the performance of estimations on available data. More precisely, 
assume we have a sample  of $n+m$ observations  $x_{-(n-1)},\dots,x_0,x_1,\dots,x_m$ at hand. We aim at $m$ backtestings, each based on a sample of length $n$: first,  we compute $m$ risk estimators  $\hat\rho_1,\ldots,\hat\rho_m$ where each risk estimator $\hat\rho_t=\hat\rho(x_{t-n},\dots,x_{t-1})$ is based on a historical sample of size $n$, starting at $t-n$ and ending at $t$. The corresponding realization of the associated P\&L is given by $x_t$ and the associated secured position (see Equation \eqref{eq:secured position}) is given by
\begin{equation}\label{eq:yt.def}
y_t:=x_t+\hat\rho_t.
\end{equation}
Furthermore, we denote by  $y:=(y_t)_{t =1,\dots,m}$ the sample for the $m$ backtestings. Sometimes, if we want to underline the dependence of $y$ on the underlying estimator $\hat\rho$, we write $y^{\hat \rho}$ instead of $y$.

\begin{remark}[Properties of $y$]\label{rem:iid}
{ Even if the initial sample $(x_t)_{t=1,\ldots,m}$ satisfies the  i.i.d.~assumption, the associated secured position $y$ will no longer be i.i.d. Indeed, if we write Equation \eqref{eq:yt.def} in full detail, i.e.
\begin{equation}
y_t=x_t+\hat\rho_t(x_{t-n},\dots,x_{t-1}),
\end{equation}  
it becomes obvious that $\hat \rho_t$ depends on $x_{t-n}$ to $x_{t-1}$, such that $y_t$ and $y_{t+1}$ depend on an overlapping sample and, hence, independence is lost in general. 
Still, the majority of the backtesting statistics ignore  this fact and assume that $\hat\rho_t$ is  equal to the true (but unknown) risk. In this case $\hat \rho$ would be constant,  hence $y_t$ would only depend on $x_t$ and so would be independent of $y_{t+1}$.

The typical approach to backtesting is the  so called  {\it traffic light} approach and we refer to Section~\ref{ex:VAR.bt} where this is discussed in reference to  regulatory VaR backtesting.} 
\end{remark}

The key to the targeted duality is the observation that value-at-risk and expected shortfall are risk measures indexed by the {\it confidence level} $\alpha$. 
More generally, consider a family of monetary risk measures indexed by a parameter $\alpha\in (0,1]$, such that the family is \emph{monotone} and \emph{law-invariant}: that is, every member of the family  is given by a distribution-based risk measure $R_\alpha: \bF \to \bR\cup \{+\infty\}$ 
and for any distribution $F\in\bF$ and $\alpha_1,\alpha_2\in (0,1]$ such that $\alpha_1<\alpha_2$, we have 
\[
R_{\alpha_1}(F)\geq R_{\alpha_2}(F).
\]
The typical task is to provide the (daily) risk estimations for some pre-defined reference risk level $\alpha_0\in (0,1]$, e.g. $0.01$ or $0.025$. The goal of backtesting is to validate the estimation methodology by quantifying the performance of the estimates using the sample of secured positions $y$.

A natural choice for the objective function in the validation procedure is the performance measure that is dual to the family $(R_\alpha)$ in the sense of \cite{CheMad2009}. Here, we concentrate on the empirical  counterpart of the dual performance measure which is obtained by replacing the unknown distribution by its empirical counterpart. More precisely, we introduce a dual empirical performance measure of the secured position $y$ in reference to the family $(R_\alpha)$ given by
\begin{equation}\label{eq:B}
	B(y):=\inf \{\alpha\in (0,1]:\, R_{\alpha}^{\textrm{emp}}(y) \leq 0\},
\end{equation}
where $R_{\alpha}^{\textrm{emp}}(y):=R_{\alpha}(\hat F_y)$ is the empirical plug-in risk estimator based on the empirical distribution  $\hat F_y$ for sample $y$ of length $m$. Intuitively speaking, in \eqref{eq:B} we look for the smallest value of $\alpha\in (0,1]$ that makes the secured sample $y$ acceptable in terms of the empirical estimators $(R^{\textrm{emp}}_\alpha)$.

\begin{remark}[Types of backtests]
{ In the literature, there is no unanimous definition of backtesting and there exist multiple performance evaluation frameworks. In this article, we follow the regulatory backtesting framework whose main aim is to assess secured position conservativeness, see e.g. \cite{Bas1996}. Note that this is essentially different from comparative backtesting based on elicitability, when the overall fit is assessed, see~\cite{Gneiting2011}. In other words, regulatory backtests are focused on risk underestimation, while comparative backtests penalize both underestimation and overestimation. Also, we focus on the unconditional coverage backtesting as the size and independence of breaches are typically assessed visually, see~\cite{McN1999} for conditional coverage tests including Christoffersen's test.}
\end{remark}

\subsection{The relation to bias}

Next, we analyze  the connection between the backtesting function introduced in \eqref{eq:B} and bias. First, the inner part of Equation \eqref{eq:B}, $R_{\alpha}^{\textrm{emp}}(y) \leq 0$, is the empirical counterpart of (risk) sufficiency. Consequently, if the risk estimator $\hat \rho$ is  sufficient, one expects $B(y)$ not to exceed the reference level $\alpha_0$ which should in turn indicate good performance of the secured position. Second, the minimum in \eqref{eq:B} is achieved for $\alpha$ for which the empirical equivalent of risk  unbiasedness is satisfied. Thus, for an unbiased estimator, the value of $B(y)$ should be close to the reference threshold value  $\alpha_0$. Consequently, one would expect that  unbiased estimators perform well in backtesting.

These two observations show that there is a direct relationship between backtesting performance and  unbiasedness. For the  computation of the empirical bias of $\hat\rho$ one could simply compute the value $R^{\textrm{emp}}_{\alpha_0}(y^{\hat \rho})$, where $\alpha_0$ is the reference risk level. Nevertheless, this  quantity is difficult to interpret as the net size of bias could vary. For instance, the net bias is not scale invariant and it is proportional to the (current) underlying position size or the current market volatility. 

To overcome this limitation,  we switch to identifying  the risk level  that makes a position acceptable (unbiased) in \eqref{eq:B}. It should be emphasized that this approach is in fact common market practice and is embedded in the regulatory VaR backtesting, see Section~\ref{ex:VAR.bt} for details. 

Summarizing, the above shows the importance of the concept of performance measures (also called {\it acceptability indices}) introduced in \cite{CheMad2009} and shows that the approach introduced therein is in fact quite generic and already adopted by practitioners; see also \cite{BieCiaDraKar2013}.

\section{The Backtesting of Value-at-risk}\label{ex:VAR.bt}

The classical regulatory  backtest for a value-at-risk (VaR) estimator (recall the definition in Equation \eqref{eq:var}) counts the empirical number of overshoots 
and compares it to the expected number of overshoots. More precisely, given the positions  $y=(y_1,\dots,y_m)$ secured by the  estimator $\hat \rho$, an  {\it overshoot} or a \emph{breach} occurs at time $t$ when the position is not sufficiently secured, i.e.~when $y_t =x_t + \hat \rho_t< 0$. Hence,  it is natural to consider the  average breach count measured by the \emph{exception rate} 
\begin{equation}\label{eq:T.def}
T(y):=\frac 1 m \sum_{t=1}^{m}\1_{\{y_t<0\}}, \qquad y \in \R^m.
\end{equation}
This measure, or  equivalently the non-standardized breach count $m T(y)$, is the standard statistic used in regulatory backtesting, see \cite{Bas2013}. 
Now, it is relatively easy to show that
\[
T(y)=\inf \{\alpha\in (0,1]:\, \var_{\alpha}^{\textrm{emp}}(y) \leq 0\},
\]
where $\var_{\alpha}^{\textrm{emp}}(y):=-y_{(\lfloor n\alpha\rfloor+1)}$ corresponds to the  empirical VaR at level $\alpha\in (0,1]$, the value $y_{(k)}$ is the $k$-th order statistic of the data, and the value $\lfloor z \rfloor$ denotes the integer part of $z\in\bR$. In other words, \eqref{eq:T.def} is a direct application of \eqref{eq:B} in the VaR context which indicates direct relation between risk bias and backtesting in the VaR framework. See also Equation (6) in \cite{BigTsa2016} where the link between risk bias and backtesting in the VaR context is outlined.

\subsection{Regulatory backtesting}
\label{sec:regulatory backtesting}
For regulatory backtesting we consider the \emph{traffic-light approach}: namely, we use yearly time series ($m=250$ historical observations) and  the estimator $\hat \rho$ is said to be in the \emph{green zone}, if $T(y^{\hat \rho}) < 0.02 $, in the \emph{yellow zone}, if $ 0.02 \le T(y^{\hat \rho}) < 0.04$, and in the \emph{red zone}, if $T(y^{\hat \rho}) \ge 0.04$. This corresponds to less than five, between five and nine, and ten or more breaches, respectively; we refer to \cite{Bas1996} for more details.

To show that bias is indeed directly reflected in backtesting we consider two examples in Section~\ref{S:var1} and Section~\ref{S:var2}, see \cite{GerTsa2011} for more examples. 
The presented results will illustrate that the usage of biased risk estimators  leads to a systematic underestimation of risk. This important defect becomes more pronounced with heavier  tails of the underlying distribution, increased confidence level (i.e.~decreasing $\alpha$),  or reduced sample size. Also, it turns out that an existing bias negatively effects the predictive accuracy as defined in \cite{Gneiting2011}. For example, it was shown  in Section 8.3 in \cite{PitSch2016} that the consistent VaR score of the Gaussian unbiased  estimator is better than the score of the standard Gaussian plug-in estimator.

\subsection{Backtesting under normality}\label{S:var1}
In this section we show that in the Gaussian setting, many popular VaR risk estimators are  biased, even in an asymptotic sense. Assume that the observed sample $x_{-(n-1)},\dots,x_m$ is an i.i.d realization from a normal distribution with unknown parameters. For simplicity, we fix $\alpha_0=1\%$, $n=250$, i.e.~backtesting on a yearly basis and consider the following four quantities:
first, recall from Example \ref{ex:Gaussian}, that   the \emph{true risk} (i.e.\ the VaR of the underlying    distribution) is given by \[
\hat \var_t^{\textrm{true}}:=-\left(\mu + \sigma\,\Phi^{-1}(0.01)\right), \qquad t=1,\dots,m.
\]
Of course, the true parameter $\theta=(\mu,\sigma)$ is only known in simulated scenarios, which we use later for testing the performance of competitive approaches. Note that having identically distributed random variables implies that true risk does not depend on $t$.

Second, if we denote the parameter estimates in backtesting period $t$ by 
$\hat \mu_t = n^{-1}\sum_{i=1}^n y_{t-i}$, and $\hat \sigma_t^2 = (n-1)^{-1} \sum_{i=1}^n (y_{t-i} - \hat \mu_t)^2$,  we obtain the common plug-in-estimator of the value-at-risk, 
\begin{equation}\label{eq:norm}
\hat \var^{\textrm{plug-in}}_t  :=-\left(\hat\mu_t +\hat\sigma_t \Phi^{-1}(0.01)\right) \qquad t=1,\dots,m.
\end{equation}

Third, as stated in Example~\ref{ex:Gaussian}, the \emph{unbiased estimator} of value-at-risk can be computed (see  \cite{PitSch2016} for details) and is given by 
\begin{equation}\label{eq:unbiased}
\hat \var^{\textrm{u}}_t  :=-\bigg(\hat\mu_t +\hat\sigma_t \sqrt{\frac{n+1}{n}}t_{n-1}^{-1}(0.01)\bigg).
\end{equation}

Fourth, it is natural to consider an empirical quantile as the estimator for value-at-risk. We call this estimator the \emph{empirical estimator} and recall that\footnote{To estimate empirical VaR we used the standard {\it quantile} function built into R software with default (type 9) setting.} 
\begin{align}\label{eq:varemp}
	 \hat \var_t^{\textrm{emp}}=-y_{(\lfloor n\alpha\rfloor+1)}.
\end{align}

\begin{figure}[tp!]
\begin{center}
\includegraphics[width=0.45\textwidth]{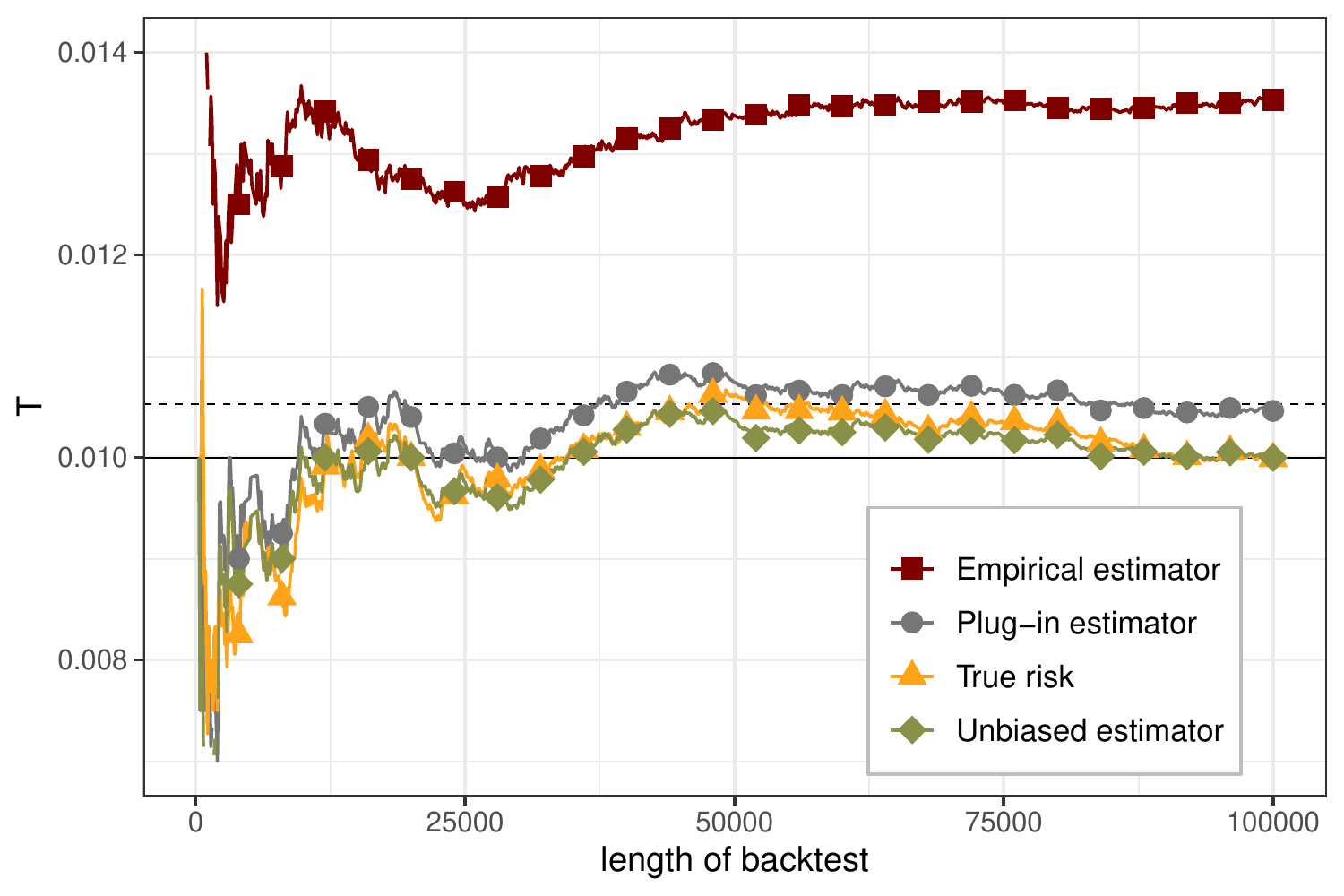}

\end{center}
\caption{VaR backtesting results under normality for increasing observation length $m$ up to  100\,000 days and a rolling window of length $n=250$.  We present the average number of exceptions $T^z$, at level  $\alpha_0=1\%$, for $\hat \var_t^{\textrm{emp}}$ (empirical), $\hat \var_t^{\textrm{plug-in}}$ (plug-in), $\hat \var_t^{\textrm{u}}$ (unbiased) and under the assumption of full knowledge of the underlying distribution, $\hat \var_t^{\textrm{true}}$ (true). Only the true (yet unknown) risk and the unbiased estimator exception rates are close to the theoretical exception rate.}
\label{F:asympt1}
\end{figure}

We are now ready to perform the backtesting and to illustrate the impact of  an existing bias on the estimator's performance. To this end, we consider a large Gaussian sample,
and construct  backtests  according to Equation \eqref{eq:T.def}: for the estimators specified by $z\in \{\textrm{true},\textrm{plug-in},\textrm{u},\textrm{emp}\}$, the average number of exceptions up to time $m$  is given by
\begin{equation}\label{eq:Tzh}
T^\textrm{z} =T^{\textrm{z}}(m):=\frac 1 m \sum_{t=1}^{m} \1_{\{x_t+\hat \var^{\textrm{z}}_t <0\}}.
\end{equation}
For a numerical illustration we keep the window length of historical data, $n$, fixed and consider $m\to \infty$. For the case of standard normal random variables (i.e.~$\mu=0$, $\sigma=1$), we plot the results in Figure~\ref{F:asympt1}.

Ideally, the value $T^\textrm{z}(m)$ should converge to theoretical reference value $\alpha_0=1\%$ when $m \to \infty$. At first sight a bit surprising, Figure~\ref{F:asympt1} shows that this is neither the case for the plug-in estimator nor the empirical estimator. 

However, for the unbiased estimator, and -- of course -- for the true risk, convergence to the theoretical reference value holds. In fact, the asymptotic exception rate for the plug-in estimator can be directly computed:  for $n=250$, 
{ the strong law of large numbers in combination with the continuous mapping theorem implies that 
\begin{align}
T^{\textrm{plug-in}}(m) \xrightarrow[m \to \infty]{a.s.} &\ 
\bE\big[  \1_{\{X+\hat \var^{\textrm{plug-in}}_t <0\}} \big] \notag\\
=&\ 
P\Big(X-\left(\hat\mu+\hat\sigma\Phi^{-1}(0.01)\right)<0\Big)\nonumber\\
 =& \ P\bigg(\frac{N-\hat\mu}{\hat\sigma\sqrt{1+1/n}}<\sqrt{\frac{n}{n+1}}\Phi^{-1}(0.01)\bigg)\nonumber\\
= & \ t_{n-1}\left(\sqrt{\frac{n}{n+1}} \Phi^{-1}(0.01)\right)\approx 1.05\%,\label{eq:T.limit}
\end{align}
where $X$ is normally distributed, $\hat\mu$  and $\hat\sigma$ the above introduced estimators of mean and  standard deviation, both based on a sample of fixed size $n$.}

The simulations also show that the asymptotic exception rate of the empirical estimator is surprisingly large and oscillates around 1.35\%. This implies serious monitoring consequences: under the correct model setting for VaR at level $\alpha=1\%$, the  breach probability in the backtest should also be 1\%. Then, the probability of reaching the yellow or the red zone  (having five or more exceptions in the annual backtest, see Section \ref{sec:regulatory backtesting}) is equal to approximately $10\%$. On the other hand, if the individual breach probability is equal to 1.35\%, the probability of reaching the yellow or the red zone is equal to approximately 25\%.\footnote{We get $1-F_{B(n,0,01)}(4)\approx 10\%$ and $1-F_{B(n,0,0135)}(4)=25\%$, where $B(n,p)$ is the Bernoulli distribution with $n=250$ trials and probability of success equal to $p=1\%$ and $p=1.35\%$, respectively.} Hence, monitoring the performance by this measure will announce problems with the model two and a half times more often as it should. Intuitively, this is due to a severe underestimation of the risk by the empirical estimator - insufficient capital leads to significantly more overshoots.



If, on the other hand, the length of  the estimation window is increased, formally referring to the case $n\to \infty$, the exception rate converges to $1\%$ also for the empirical and the plug-in estimator, as expected.

\begin{remark}[Link to predictive inference methods]{
The proposed approach leading to the unbiasedness of risk estimators detailed in  Equation \eqref{eq:rho.unbiased} follows the principle of predictive inference propagated amongst others in \cite{billheimer2019predictive}. Instead of focusing on the estimation of parameters, the goal is to predict future observations, or, more precisely, their inherent risk. As in the case of the {\it predictive confidence interval}, see \cite[Example 2.2]{Gei1993}, this implies the change from the Gaussian distribution to the $t$-distribution, which can be spotted in Equations \eqref{eq:norm} and \eqref{eq:unbiased}.

In other words, the unbiased estimator is taking into account the fact that estimated capital will be used to secure a future and risky position and requires that the estimation procedure acknowledges parameter uncertainty.

It seems interesting to note  that  predictive inference methods can be applied to get unbiased estimators in other settings, i.e. for other distributional families and related estimation techniques. For example, under the \emph{Pareto} distribution, one can show that while the standard VaR plug-in estimator $-\hat\mu \log(1-p)$ is (risk) biased, the adjusted estimator $-\hat\mu \log(1-p^*)$, with $$p^*:=1-\exp\{-n[(1-p)^{-1/n}-1]\}$$ is  unbiased.
For more details, we refer to \cite{GerTsa2011}, where the link between failure probability and uncertainty embedded into parameter estimation is discussed.}
\end{remark}

\subsection{Backtesting under generalized Pareto distributions (GPD)}\label{S:var2}

In the context of heavy tails, a closed-form expression for the unbiased estimator is  no longer available and one has to rely on numerical procedures. We introduce such an approach in the following section. We begin by describing the currently existing estimators for generalized Pareto distributions and illustrate their performance.

Consider the i.i.d.~setting from Section \ref{S:2} and  assume that $x_{-(n-1)},\dots,x_0,x_1,\dots,x_m$ follows a GPD  (in the left tail) with threshold $u\in\bR$, shape $\xi\in \bR\setminus \{0\}$, and scale $\beta > 0$; we also set $p:=P(X\le u)$. In practical applications, the threshold $u$ is often considered as known while the parameters $\xi$ and $\beta$ have to be estimated.

Under the assumption of a fixed and known (left tail) distribution it is well-known how to compute the value-at-risk under GPD, we refer to \cite{MFE} for details. Indeed, basic calculations yield the true risk value
	\begin{align*}
		\var_t^{\textrm{true}}:=-u+\tfrac{\beta}{\xi}\Big(\tilde \alpha^{-\xi}-1\Big),
	\end{align*}
where $\tilde\alpha:=\alpha/p$ is threshold-adjusted confidence level. Denoting by $\hat \xi_t$ and $\hat \beta_t$ (with additional $t$ referring to the estimation period) the probability weighted moments (PWM) estimators for $\xi$ and $\beta$ we readily obtain the plug-in estimator for value-at-risk in the GPD case\footnote{One could alternatively use different estimation technique to obtain the plug-in estimator, based e.g. on the Maximum Likelihood framework; for most plug-in procedures (that do not take into account predictive inference), the conclusions from this section should apply.}, 
	\begin{align*}
		\hat \var_t^{\textrm{plug-in}}:=-u+\tfrac{\hat\beta_t}{\hat\xi_t}\Big(\tilde\alpha^{-\hat\xi_t}-1\Big).
	\end{align*}
When the threshold $u$ needs to be estimated, we would additionally replace $u$ by $\hat u_t$. Moreover, to numerically asses an existing bias in the estimation, we also consider the empirical estimator $\hat \var_t^{\textrm{emp}}$, already introduced previously in Equation \eqref{eq:varemp}. 

\begin{figure}[tp!]
\begin{center}
\includegraphics[width=0.45\textwidth]{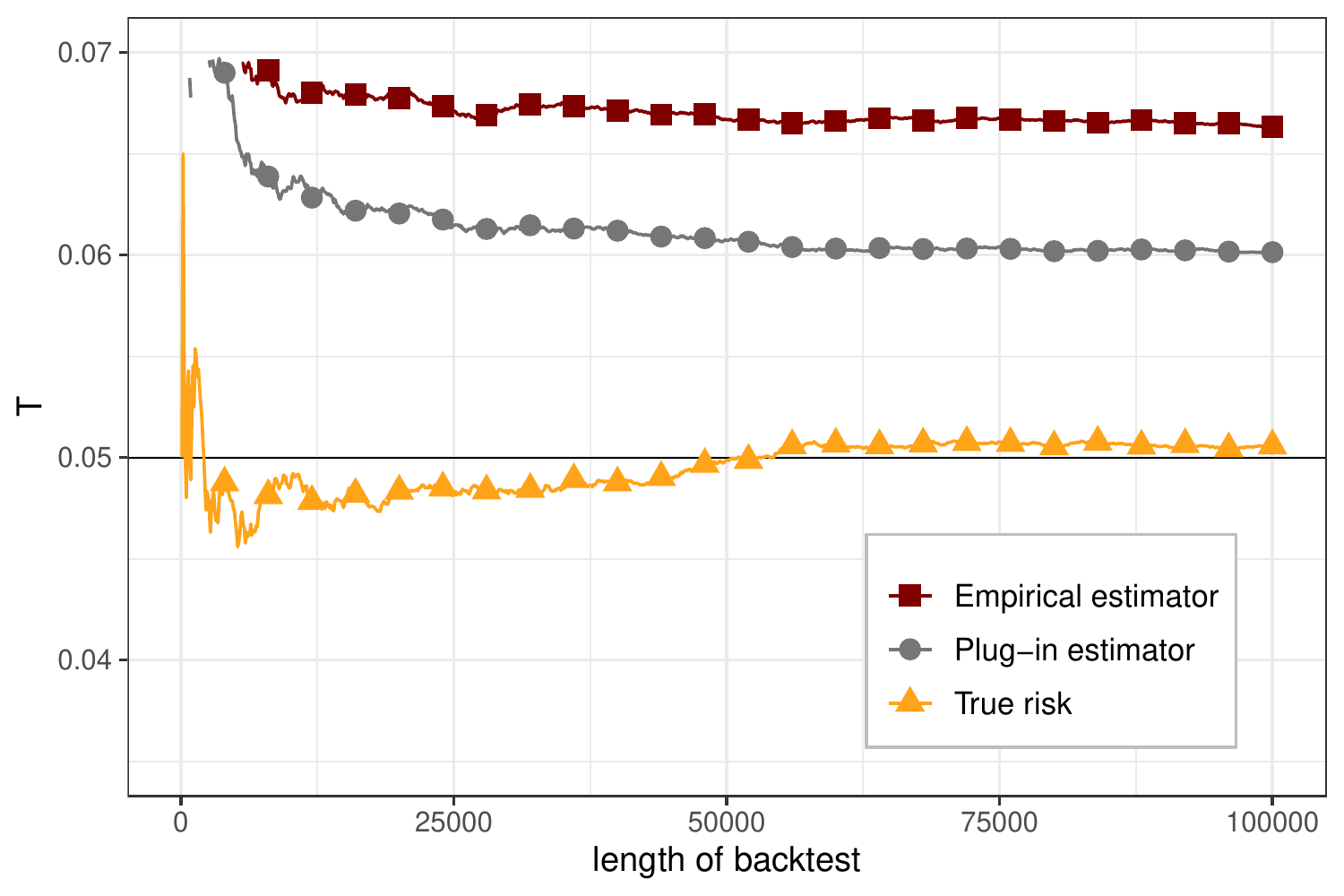}
\end{center}
\caption{VaR backtesting results under GPD for increasing observation length $m$ up to 100\,000 days and a rolling window of length $n=250$. We present the average number of exceptions $T^z$, at level  $\alpha_0=1\%$,  for $\hat \var_t^{\textrm{emp}}$ (empirical), $\hat \var_t^{\textrm{plug-in}}$ (plug-in),  and under the assumption of full knowledge of the underlying distribution, $\hat \var_t^{\textrm{true}}$ (true). Only the true (and unknown) risk exception rate is close to the theoretical exception rate.}
\label{F:asympt2}
\end{figure}

For simplicity, we consider a fixed parameter set  $u=-1$, $\xi= 0.05$, $\beta = 0.7$, $p=0.2$ and assume conditional sampling. Since only 20\% of the data lie beyond the threshold $u$ and the observed sample consists only of the data below the threshold, the corresponding periods have to be adjusted accordingly. We consider a learning period of length $n=250 \cdot p = 50$ and adjusted reference level $\tilde \alpha_0=0.01/p=0.05$. For $z\in \{\textrm{true},\textrm{plug-in},\textrm{emp}\}$,  we construct the secured position and perform the backtest according to Equation  \eqref{eq:Tzh}. Again, note that for simplicity we have assumed that we are given a consistent set of rolling left-tail observations and ignored non left-tail (above the threshold) inputs. The results are shown in Figure~\ref{F:asympt2}.

The figure considers a fixed window size, $n=50$, together with increasing sample size, i.e.~$m\to \infty$. We observe that the bias vanishes for the true value-at-risk -- as expected there is no bias once the true distribution is known. On the contrary, both the plug-in estimator and the empirical estimator show a clear bias. The asymptotic risk level reached by $\hat \var_t^{\textrm{plug-in}}$ is around 6\% (instead of 5\%), while for $\hat \var_t^{\textrm{emp}}$ it is even close to 7\%. The latter means that in 7\% of the observed cases the estimated risk capital is not sufficient to cover the occurred losses which corresponds to a significant underestimation of the present risk.

\section{Expected Shortfall backtesting}\label{ex:ES.bt}
 In this section we show that our observations on bad performance of biased estimators for VaR also hold true for expected shortfall (ES).  ES is a well-recognised measure in both the financial and the  insurance industry. For a financial position $Z$ with value-at-risk $(\var_{\alpha}(Z))_{\alpha \in (0,1]}$ the expected shortfall at level $\alpha\in (0,1]$ is given by
\begin{equation}\label{eq:es}
\ES_{\alpha}(Z):=\frac{1}{\alpha}\int_0^{\alpha}\var_{\gamma}(Z)d\gamma.
\end{equation}
Since value-at-risk is law-invariant, so is expected shortfall and the associated representation can immediately be obtained from \eqref{eq:var}.

Next, we illustrate the backtesting performance in case of expected shortfall by mimicking the framework introduced in Section~\ref{ex:VAR.bt}. The duality-based performance metric based on \eqref{eq:B} can also be  used for ES backtesting. For the sample $y$ secured by ES,  the empirical mean of the overshooting samples is given by
\begin{equation}\label{eq:G.def}
G(y):=\frac{1}{m}\sum_{t=1}^{m}\1_{\{y_{(1)}+\ldots+y_{(t)}<0\}},
\end{equation}
where $y_{(k)}$ denotes the $k$th order statistic of $y$.
The performance statistic \eqref{eq:G.def} simply measures the cumulative breach count and answers a simple question: how many (worst-case) scenarios do we need to consider to know that the aggregated loss does not exceed the aggregated capital reserve, see~\cite{MolPit2017} for details. As expected, \eqref{eq:G.def} is the empirical counterpart of \eqref{eq:B}, i.e. we have
\[
G(y)=\inf \{\alpha\in (0,1]:\, \ES_{\alpha}^{\textrm{emp}}(y) \leq 0\},
\]
where $\ES^{\textrm{emp}}_{\alpha}$ denotes the empirical expected shortfall at level $\alpha\in (0,1]$ estimator. As in Section~\ref{S:var1} and Section~\ref{S:var2}, we show backtesting results for Gaussian and GPD distributions. In both cases, we replace average exception rate statistic with \eqref{eq:G.def}. Hence, for every estimator $z$ in scope, we consider
\begin{align}\label{eq:G}
G^z=G^z(m):=\frac{1}{m}\sum_{t=1}^{m}1_{\{y^z_{(1)}+\ldots+y^z_{(t)}<0\}},
\end{align}
where, for each $m$, the order statistic $y^z_{(k)}$ corresponds to the $k$th order statistic of the associated secured position  
$$y^z_t = x_t+\hat \ES^{\textrm{z}}_t, \qquad t=1,\dots,m, $$ and where  $\ES^{\textrm{z}}_t$ denotes the $t$-th day estimated expected shortfall risk using estimator $z$. 

\subsection{The Gaussian setting}Under the Gaussian i.i.d.\ setting  introduced in Section \ref{S:var1}, we consider the same set of estimators, now defined for reference level $\alpha_0=2.5\%$, with the same learning sample size  as before, i.e.\ $n=250$. In this setting, we consider {\it true risk}, {\it Gaussian plug-in}, {\it Gaussian unbiased}, and {\it empirical} estimators given by
\begin{align*}
	\ES^{\textrm{true}}_t   & := -\mu + \sigma \frac{\phi(\Phi^{-1}(\alpha_0))}{\alpha_0}, \\
	 \hat \ES^{\textrm{plug-in}}_t & :=-\hat\mu_t +\hat\sigma_t \frac{\phi(\Phi^{-1}(\alpha_0))}{\alpha_0},\\
	\hat \ES^{\textrm{u}}_t  & :=-\hat\mu_t +c_{250} \cdot \hat\sigma_t \frac{\phi(\Phi^{-1}(\alpha_0))}{\alpha_0}, \\
	\hat \ES_{t}^{\textrm{emp}} &:=-\left(\frac{ \sum_{i=t-n}^{t-1}x_i\1_{\{x_i+\hat{\var}_{t}^{\textrm{emp}}\leq 0\}}}{\sum_{i=t-n}^{t-1}\1_{\{x_i+\hat{\var}_{t}^{\textrm{emp}}\leq 0\}}}\right),
\end{align*}
where the constant $c_{250}=1.0077$ is obtained using approximation scheme introduced in Example 5.4 in \cite{PitSch2016}. 

\subsection{The GPD setting}Similarly, in the GPD i.i.d.\ setting  introduced in Section \ref{S:var2},  we consider the {\it empirical} estimator as well as the {\it true risk} and the {\it plug-in} estimators given by
\begin{align*}
\ES_t^{\textrm{true}}&:=\frac{\var_t^{\textrm{true}}}{1-\xi} +\frac{\beta+\xi u}{1-\xi}, \\
\hat \ES_t^{\textrm{plug-in}}&:=\frac{\hat \var_t^{\textrm{plug-in}}}{1-\hat\xi_t} +\frac{\hat\beta_t+\hat\xi_t u}{1-\hat\xi_t},
\end{align*}
for conditional reference risk level $0.025/p=0.125$, and conditioned sample size $250 p=50$, see \cite{MFE} for details. Note that we need to assume $\xi<1$ for expected shortfall to be finite.

\begin{figure}[tp!]
\begin{center}
\includegraphics[width=0.45\textwidth]{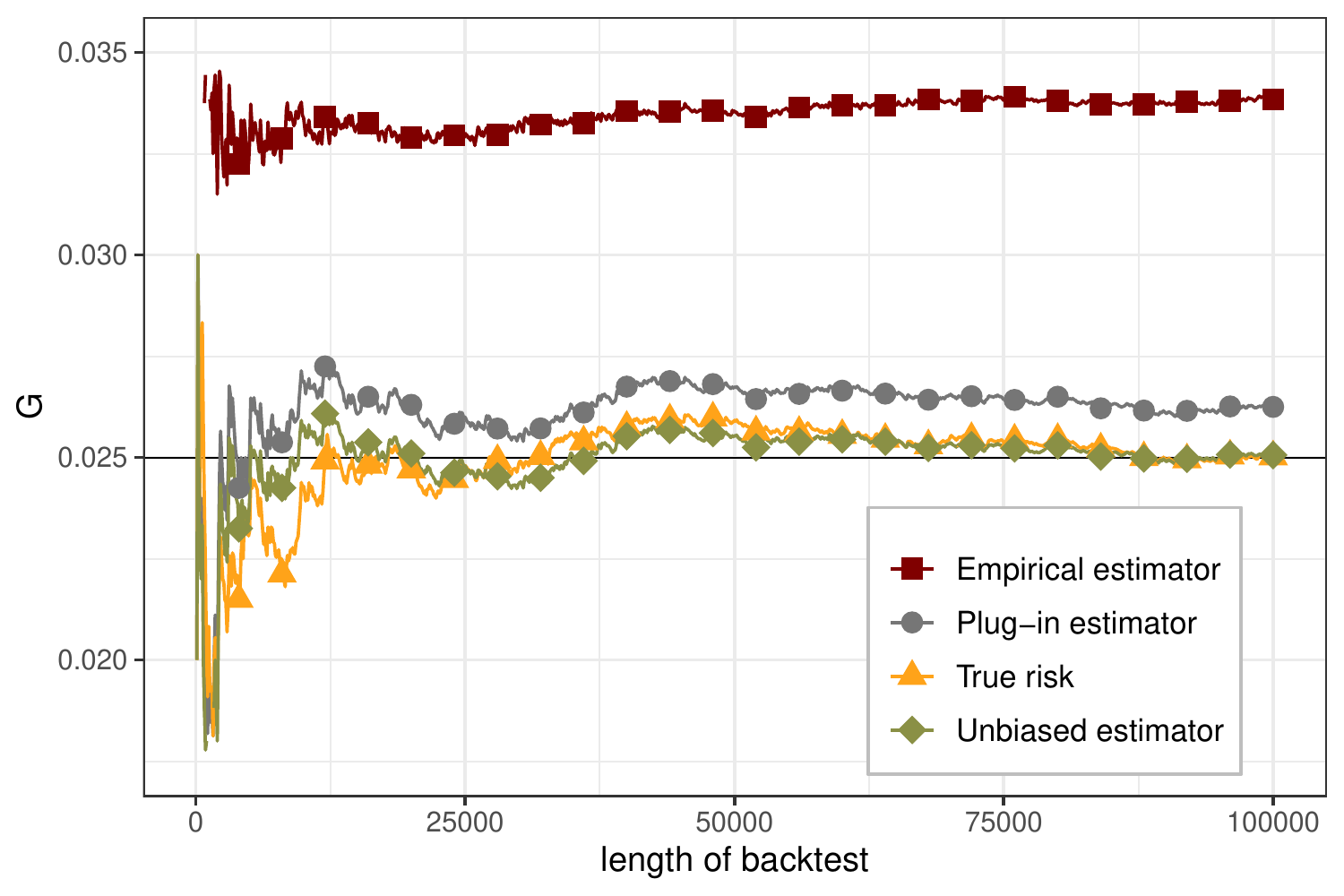}
\includegraphics[width=0.45\textwidth]{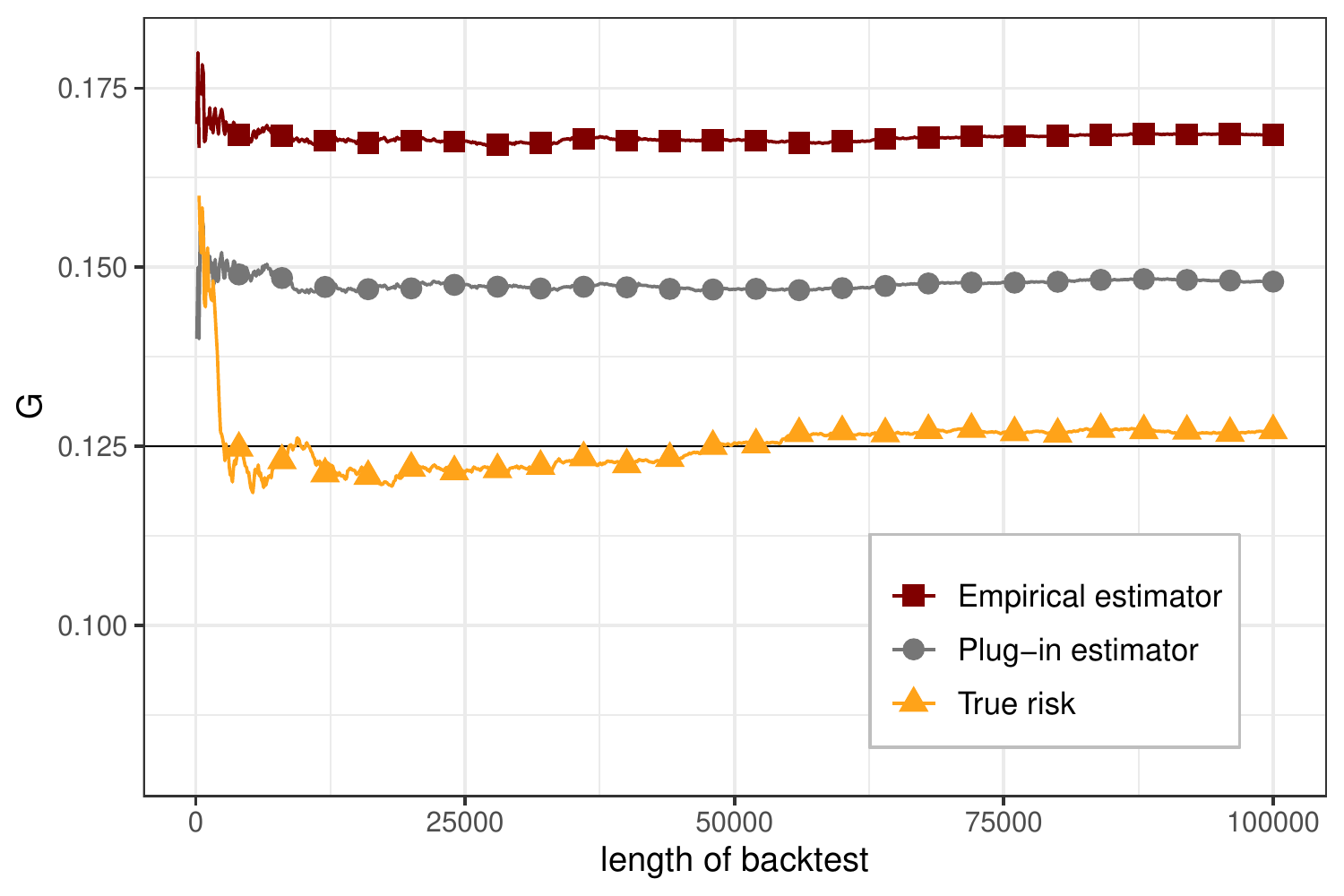}
\end{center}
\caption{ES backtesting results under normality (left) and GPD (right) for increasing observation length $m$ up to  100\,000 days and unconditional rolling window of length $250$ and $50$, respectively. Empirical mean of the overshoots $G^z$ at level 2.5\% and 12.5\%, respectively, is presented for selected estimators. Only the true (yet unknown) risk and the unbiased estimator rates are close to the theoretical rates.}
\label{F:asympt1b}
\end{figure}
In both cases we consider the same set of parameters that was used in Section~\ref{S:var1} and Section~\ref{S:var2}, respectively, and plot values of backtesting statistic $G^z$. The aggregated results for both normal case and GPD case are presented in Figure~\ref{F:asympt1b}. As in the value-at-risk case,  only the estimator who knows the underlying distribution (true risk) and the Gaussian unbiased estimator show no bias. In all other cases, a significant bias is visible.

\section{Backtesting in a general context}\label{S:new.one}
While up to now we focussed mainly on value-at-risk and expected shortfall,  the link between estimation procedure and backtesting results turns out to be fairly general and can be established for a large class of risk measures. We illustrate this by considering risk measures based on expectiles, see Section~\ref{S:evar}, a family of risk measures which recently gained a lot of attraction  due to the fact that it combines coherence with elicitability, see \cite{NolZie2017} for details.

In addition we consider backtesting beyond the i.i.d.~case, i.e.~extend the considered framework to the heteroscedastic case. For heteroscedastic data, the link between estimation and backtesting performance  becomes even more pronounced, especially when the estimation technique is not linked directly to the underlying process dynamics. We provide a high-level illustration in Section~\ref{S:garch} using a GARCH(1,1) setting.

\subsection{Backtesting Expectile Value-at-Risk}\label{S:evar}
For a financial position $Z$, its expectile value-at-risk (EVaR) at level $\alpha\in (0,1)$ is given by
\[
\evar_{\alpha}(Z):=-\argmin_{x\in\bR}\left( \alpha \bE[(X-x)^2_{+}]+(1-\alpha)\bE[(X-x)^2_{-}] \right).
\]
Following \cite{BelDib2017}, we set the reference threshold $\alpha$ to $\alpha_{0}=0.00145$. For EVaR backtesting, given a secured position $y$, we use the distorted Gain-Loss empirical ratio 
\[
H_m(y):=h\left(\frac{E_{\textrm{emp}}[y_{+}]}{E_{\textrm{emp}}[y_{-}]}\right)=h\left(\frac{\sum_{t=1}^{m}\1_{\{y_t\geq 0\}}y_t}{-\sum_{t=1}^{m}\1_{\{y_t\leq 0\}}y_t}\right),
\]
where $h(z)=1/(z+1)$ is the distortion function and $E_{\textrm{emp}}$ refers to the expectation under the empirical distribution, which immediately implies the right hand side. Note that while the non-distorted gain-loss ratio was suggested in \cite[Section 4]{BelDib2017} as the expectile performance evaluation metric, we decided to apply a distortion to be consistent with the framework introduced in \eqref{eq:B}. 

In particular, it is natural to invert the initial ratio as it might be non-stable due to a small population of negative values in $y$. More explicitly, we set
\[
H_m(y)=\inf\left\{\alpha\in (0,1)\colon \frac{E_{\textrm{emp}}[y_{+}]}{E_{\textrm{emp}}[y_{-}]}\geq \frac{1-\alpha}{\alpha}\right\},
\]
so $H_m(y)$ is looking for a minimal threshold $\alpha\in (0,1)$ for which $y$ is falling into the empirical acceptance set of measure $\evar_{\alpha}$, cf. Equation 4 in \cite{BelDib2017} where acceptance sets for EVaR are discussed. 

Hence, $H_m$ is a special example of empirical performance measure introduced in \eqref{eq:B}. To illustrate the link between estimation bias and backtesting for EVaR let us mimic the framework introduced in Section~\ref{ex:VAR.bt} for simulated Gaussian and Student $t$  data\footnote{For simplicity, we decided to use the Student-$t$ distribution instead of the GPD distribution used in the previous paragraph. The reason for this is that since EVaR is not a conditional left tail risk measure, one would need to consider also the distribution above the threshold in the GPD setting.}.

To this end, we pick a large sample of i.i.d.~normally distributed random variables and a sample of Student-$t$ distributed ones with 5 degrees of freedom $t_5$, set the learning period to $n=250$ and perform the backtest for {\it true risks}, {\it plug-in estimators}, and {\it empirical expectile estimators}. The values for true risks, $\evar^{\textrm{true}}_t$,  were obtained using the R functions {\it enorm} and ${\it et}$ from {\it expectreg}. The plug-in estimator values, $\evar^{\textrm{norm}}_t$,  were obtained as in the VaR case, i.e. we estimated underlying parameters (mean, scale, shape)  and plugged them in into true risk parametric formulas. The empirical estimator values, $\evar^{\textrm{emp}}_t$,  were obtained using the R function {\it expectile} from {\it expectreg}; the estimation is based on LAWS procedure. 

Again, for every estimator $z$ in scope, we considered
\begin{align}\label{eq:H}
H^z=H^z(m)=(H(y_t^z))_{t=1,\ldots,m}.
\end{align}

\begin{figure}[tp!]
\begin{center}
\includegraphics[width=0.45\textwidth]{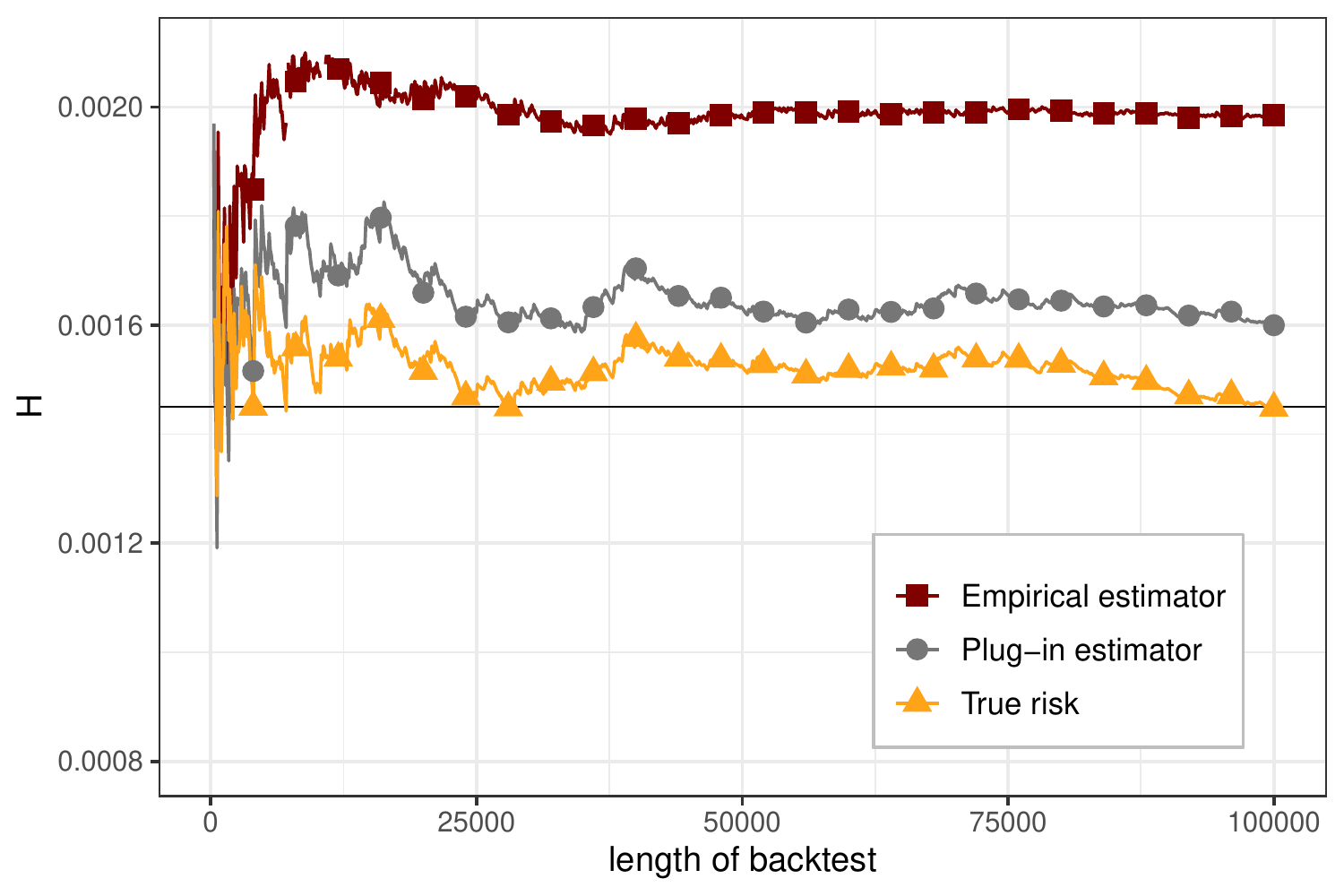}
\includegraphics[width=0.45\textwidth]{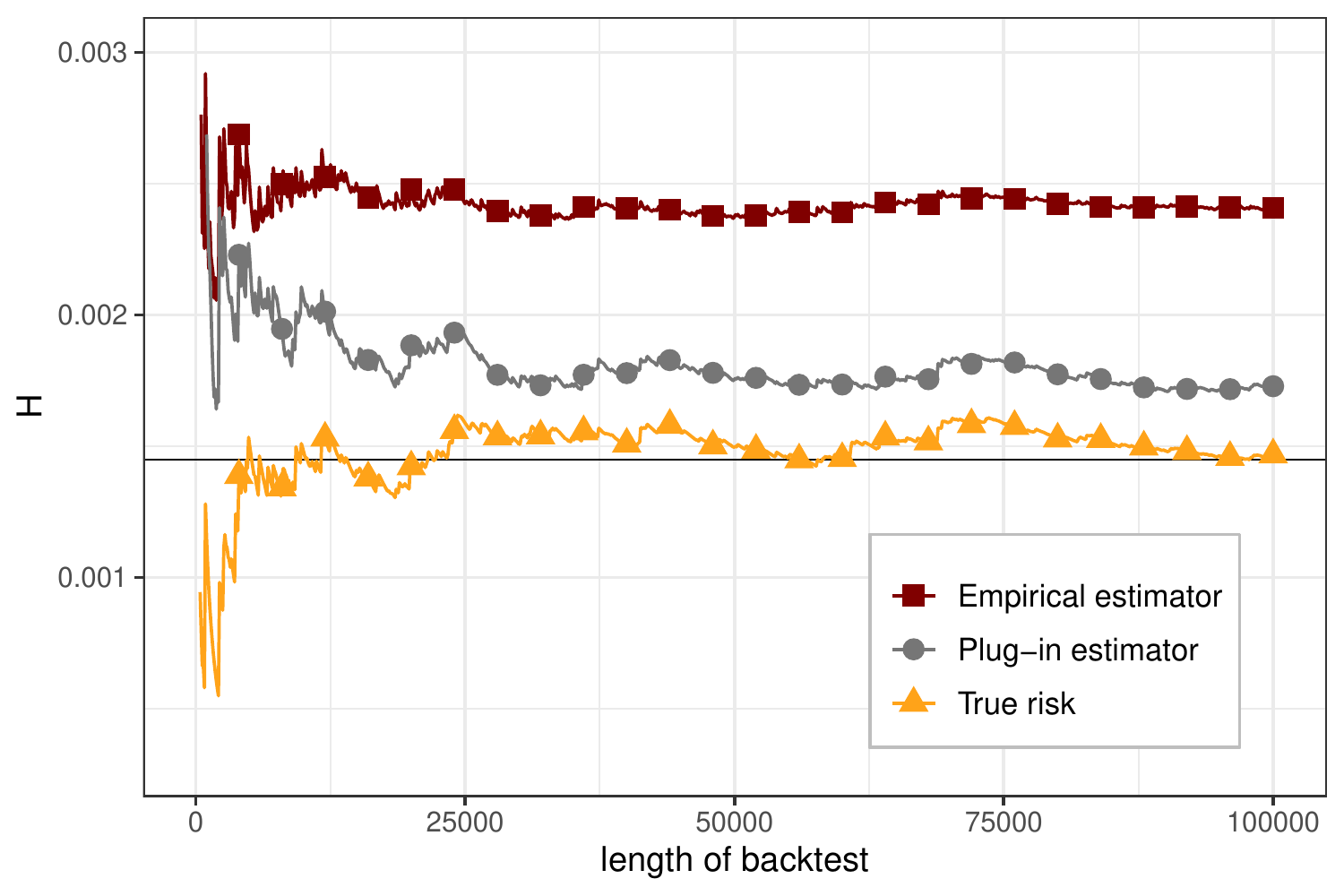}
\end{center}
\caption{EVaR backtesting results under normality (left) and Student t (right) for increasing observation length $m$ up to  100\,000 days and unconditional rolling window of length $250$. Empirical distorted gain-loss ratio $H^z$ for level 0.145\% is presented for selected estimators. Only the true (yet unknown) EVaR risk is close to the theoretical level.}
\label{F:expectile.asympt1}
\end{figure}

The results for both normal and Student-$t$ data are presented in Figure~\ref{F:expectile.asympt1}.
Again, we see a behaviour consistent with the one observed for VaR and ES. No systemic bias is observed only for the true risk estimator, which suggests that the bias presence is in fact linked to estimation procedure framework rather than specific risk measure choice.

\subsection{Backtesting in the GARCH(1,1) setting}\label{S:garch}
In this section we adopt the  framework introduced in Section~\ref{ex:VAR.bt} to the heteroscedastic case. In this regard, assume that the observed sample  of $n+m$ observations denoted by $x_{-(n-1)},\dots,x_0,x_1,\dots,x_m$ is a realisation from a GARCH(1,1) process $X=(X_t)_{t=-(n-1),\dots,m}$ following the  dynamic
\begin{eqnarray}\label{eq-1}
\begin{cases}
X_t &= \mu+\sigma_t \epsilon_t , \\
\sigma_t &= \sqrt{\omega+ \alpha X_{t - 1}^2 + \beta\sigma^2_{t - 1}}
\end{cases},\quad t=-(n-1),\ldots,m,
\end{eqnarray}
where $(\epsilon_t)$ is  Gaussian white noise, $\mu=0$, $\omega=0.0001$, $\alpha=0.1$, and $\beta=0.8$. For simplicity, we assume we are given a fixed initial (conditional) standard deviation that is equal to $\sigma_{-(n-1)}:=\omega/(1-\alpha-\beta)=0.01$.

The true conditional value-at-risk is obtained by assuming full knowledge of the underlying  parameters. Assume we have observed the process until time $t-1$, the \emph{true (conditional) value-at-risk} for $X_{t}$ is given by
\begin{equation}\label{eq:true.garch}
\var_{t}^{\textrm{true}}=-\big(\mu+ \sigma_t\Phi^{-1}(\alpha)\big) =- \sigma_t\Phi^{-1}(\alpha) 
\end{equation}
since $\mu=0$.
Note that here, $\sigma_t=\sigma_t(\omega,\alpha,\beta,X_{t-1},\sigma_{t-1})$, such that the true value-at-risk indeed depends on the parameter and the history of the process.

On the other hand,  the associated plug-in  estimator is obtained in two steps: first, the parameters  are estimated by the quasi-least squares estimators $(\hat\omega,\hat\alpha,\hat\beta)$; see Section II.4.2 in \cite{Car2009b}. Next,  $\hat \sigma_t$ is obtained recursively via \eqref{eq-1} applied to $n$ past observations with the estimated parameters and $\hat\mu_t$ is the estimated process mean. The plug-in estimator then has the form 
\begin{equation}\label{eq:plugin.garch}
\hat \var_{t}^{\textrm{plug-in}}=-\left(\hat\mu_t+ \hat\sigma_t\Phi^{-1}(\alpha)\right).
\end{equation}

For reference purposes, we also calculate the standard empirical VaR estimator, and report its values. 
To the best of our knowledge, an unbiased estimator is not available in the GARCH(1,1) context which motivates a numerical procedure for bias reduction which we introduce in the following section.

\begin{figure}[tp!]
\begin{center}
\includegraphics[width=0.45\textwidth]{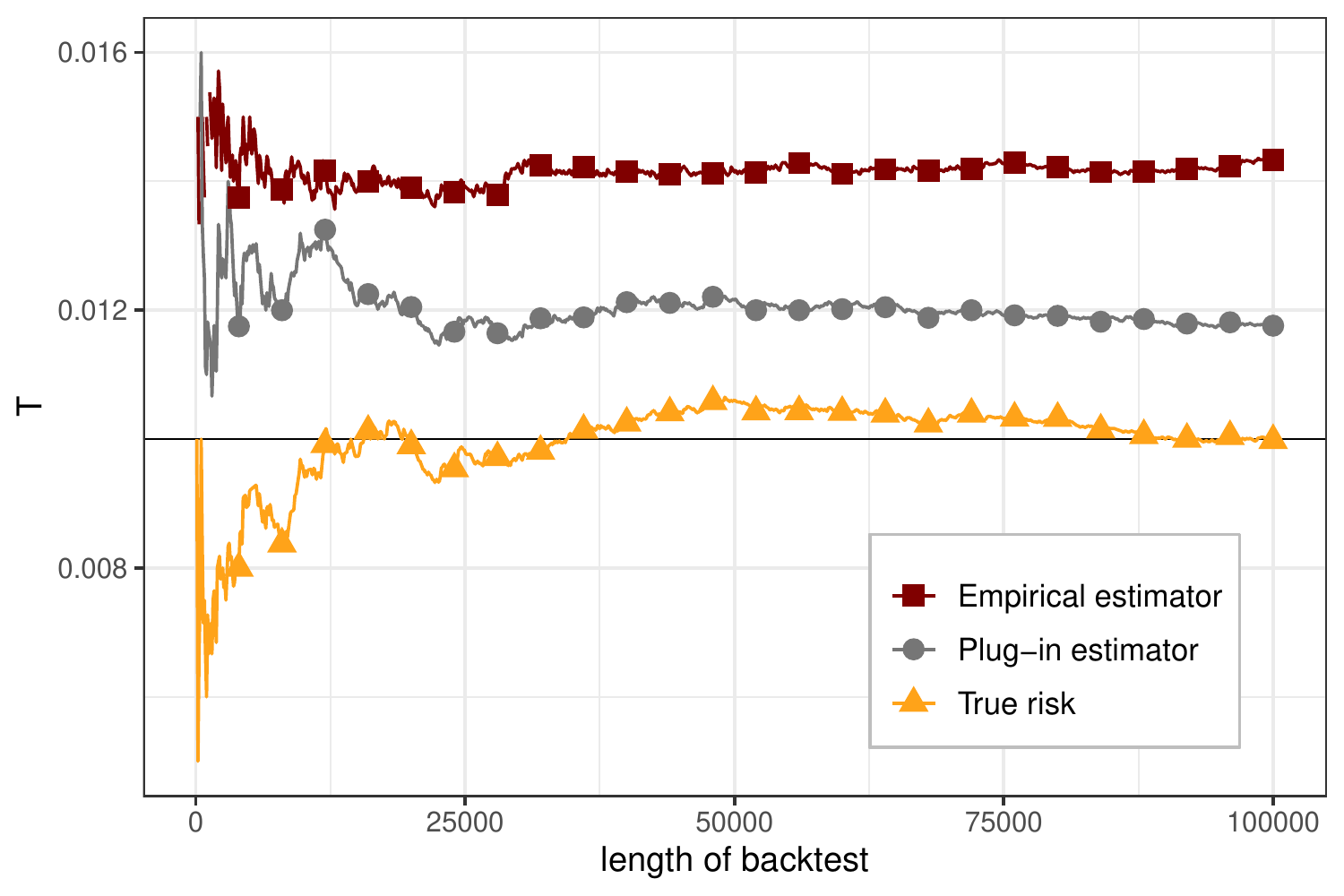}
\end{center}
\caption{VaR backtesting results under GARCH(1,1) dynamics for increasing observation length $m$ up to  100\,000 days and a rolling window of length $n=250$.  We present the average number of exceptions $T^z$, at level  $\alpha_0=1\%$, for $\hat \var_t^{\textrm{emp}}$ (empirical), $\hat \var_t^{\textrm{plug-in}}$ (plug-in) and under the assumption of full knowledge of the underlying distribution, $\hat \var_t^{\textrm{true}}$ (true). Only the true (yet unknown) risk estimator exception rates are close to the theoretical exception rate.
}

\label{F:garch.asympt1}
\end{figure}

The values of $T^z(m)$ for $z\in \{\textrm{true},\textrm{plug-in},\textrm{emp}\}$ are presented in Figure~\ref{F:garch.asympt1}.
%
%
As expected, a positive bias could be identified for both empirical and the plug-in estimator. It should be noted that due to the non-i.i.d.\ nature of the data,  the asymptotic exception rate for the empirical estimator  increases in comparison to the Gaussian setting from around 1.35\% to around 1.45\%. Also, note that the asymptotic exception rate for the plug-in  estimator (1.2\%) is bigger in comparison to the rate for the Gaussian plug-in estimator (1.05\%). Hence, we conclude that for the heteroscedastic setting considered here, the (risk) bias  increases due to additional uncertainty encoded in the dependency structure.

\section{Bias reduction for plug-in estimators}\label{S:bias.reduction}
Our previous findings show that plug-in estimators typically come with a bias implying a negative impact on their backtesting performance. However, in many situations they are the natural starting point for more elaborate estimators, which we will detail now. 

For a law-invariant risk measure, the plug-in procedure can be formalized as follows: 
assume that  the distribution of $X$ lies in the parametric family $F_\theta(X): \theta \in \Theta$ and consider a distribution-based risk measure $R:\bF \to \bR\cup \{+\infty\}$, see Section~\ref{sec:2}.  Then,  the \emph{plug-in estimator} is obtained using the following two steps: first, we estimate the parameter $\hat\theta$. Second, we plug the estimator $\hat\theta$  into the  formula for the risk measure and obtain 
\[
\hat\rho^{\textrm{plug-in}}:=R(F_{\hat\theta}(X)).
\]

Since we expect this estimator to be biased, we introduce a general scheme for improving the efficiency of an existing parametric plug-in estimator. 
As a motivation, let us revisit the Gaussian case: from Equations  \eqref{eq:unbiased} and \eqref{eq:norm}
it may be observed that the unbiased estimator is obtained from the plug-in estimator by changing the parameters according to the mapping
\begin{equation}\label{eq:gaussian.map}
(\hat \mu, \hat \sigma)\longmapsto \left(\hat \mu, \sqrt{\frac{n+1}{n}}\frac{t_{n-1}^{-1}(\alpha)}{\Phi^{-1}(\alpha)} \, \hat \sigma\right),
\end{equation}
where $n\in\bN$ is the underlying sample size. In particular, only a rescaling with respect to the underlying variance takes place and no rescaling with respect to the mean. 

This inspires the following procedure for bias-reduction: denote the $d$-dimensional estimated parameter by $\hat \theta$. We search for a transformation $a$, such that  $a \circ \hat\theta$ minimizes the bias. In most cases a linear mapping will suffice, and - as noted above - it may also be optimal to change only a few parameters and not all. 

In other words, assuming that $\Theta\subseteq \bR^d$ we aim at finding $a\in\bR^d$ for which the tweaked plug-in estimator
\[
\hat\rho_a:=R(F_{a\circ \hat\theta}(X))
\]
has the smallest bias; here $\circ$ denotes component-wise multiplication. While a global minimizer might cease to exist, we aim at a local minimization which we introduce now.

\subsection{Local bias minimization} 
In the local bias minimization we consider only a subset $\Theta_0\subseteq \Theta$ which is chosen dependent on the data at hand. 
For the distribution-based risk measure $R$ and some vector $a\in \R^d$ we define the \emph{maximal risk bias} of the estimator $\hat \rho_{a}$ on $\Theta_0$ by
\begin{equation}\label{eq:bias.B}
B_*(\hat \rho_a,\Theta_0) := \sup_{\theta \in \Theta_0} R(F_\theta(X+\hat\rho_a)).
\end{equation}

In particular, if $B_*(\hat \rho_a,\Theta)=0$, then $\hat \rho_a$ is risk sufficient. Indeed, under this condition
$$ R(F_\theta(X+\hat\rho_a)) \le B_*(\hat \rho_a,\Theta)=0 $$  
for all $\theta \in \Theta$ and hence $\hat \rho_a$ satisfies \eqref{eq:sufficient}. 

The \emph{local bias minimizer} $a_{\Theta_0}^*$ is the parameter which satisfies 
\begin{equation}\label{eq:risk.bias.est}
    a_{\Theta_0}^* \in \argmin_{a\in\bR^{d}} \big| B_*(\hat \rho_{a};\Theta_0)\big|
\end{equation}
and the estimator
\begin{align}
    \hat\rho^{\textrm{lm}}=R(F_{a_{\Theta_0}^* \circ \hat \theta}(X))
\end{align}
is the locally risk minimizing estimator. In practice, it will of course not be easy to obtain this estimator and we detail in the following a bootstrap procedure to compute an
approximation of $\hat\rho^{\textrm{lm}}$.

For completeness, we refer to \cite{BigTsa2016}, where a similar approach based on absolute risk bias adjustment has been proposed. However, note that the approach proposed in the aforementioned paper is different from ours as it aims to estimate the bias value $B_*(\hat\rho; \theta)$ directly using the bootstrap method; the estimated value is later added to estimated risk to reduce the bias. That saying, for simplified scale-location families those two approaches could produce consistent results. In particular, we note that while in \cite{BigTsa2016} the GPD case is not studied, the authors consider the Student $t$-distribution framework and show how to minimize bias in this setting. While the numerical analysis in the cited paper is not directly linked to backtesting or predictive accuracy analysis as in our case (see Section~\ref{sec:4}), absolute bias reduction oriented studies also indicate improved performance when imposing control on the underlying risk bias (called {\it residual risk} therein) e.g.\ by considering residual add-on or shifting risk measure reference risk level.

\subsection{The bootstrapping bias reduction}\label{S:boostrap.alg}

The most likely parameter $\theta$ after estimation is $\hat \theta$ and we will consider $\Theta_0={\hat \theta}$ in the following. For simplicity we write $B_*(\hat \rho_a,\theta)=B_*(\hat \rho_a,\{\theta\})$ and similar for $a^*$. 

We focus on the i.i.d.\ situation here, where 
$$ F_\theta(X+\hat\rho_a) = F_\theta(X) * F_\theta(\hat \rho_a) $$
since $X$ will be independent of the past data, and where $*$ denotes the convolution of two distributions.

The estimation of $a^*$ will be done on simulated data according to the distribution $F_{\hat \theta}$, which is called \emph{bootstrap}. The obtained optimal parameter is denoted by $a^{\textrm b}$ and the associated estimator is denoted by 
\[
\hat \rho^{\textrm{b}}:=R(F_{a^{\textrm b} \circ \hat \theta}(X)).
\]
The proposed algorithm  is summarized in details in Figure~\ref{f:boot.algorithm}.




%
\begin{figure}[tp!]
\scalebox{0.8}{
\begin{tcolorbox}
{\bf Algorithm 1: The bootstrapped risk estimator.}

\noindent Perform the estimation:
\begin{enumerate}
\item[(1)] Estimate $\hat{\theta}$ (e.g. using MLE approach) from $X_1,\dots,X_n$.
\end{enumerate}
Perform the bootstrap:
\begin{enumerate}
\item[(2)] For each $i=1,\dots,B,$ simulate an i.i.d.~sample $\xi^i=(\xi_1^i,\dots,\xi_n^i)$ of size $n$ from the estimated distribution $F_{\hat \theta}(X)$ and compute the estimator $\hat{\theta}^i$ for each $\xi^i$.
\item[(3)] For any $a\in\cA$ (where $\cA=\bR^d$ or $\cA$ corresponds to a suitably chosen subset) do the following:
\begin{itemize}
\item[(3a)] estimate the distribution  $F_{\hat\theta}(\hat \rho_{a})$ directly from the sample $(\hat \theta^1,\dots,\hat \theta^B)$, e.g.~by a kernel density estimation.
\item[(3b)] Compute the convolution $F_{\hat \theta}(X) \, * \, F_{\hat\theta}(\hat \rho_{a})$.
\end{itemize}
\end{enumerate}
Finally, compute the optimal choice of the parameter $a$:
\begin{enumerate}
\item[(4)] Calculate $a^{\textrm b}:=\argmin_{a\in\cA}\left| R\big(F_{\hat \theta}(X) \, * \, F_{\hat\theta}(\hat \rho_{a}) \big)\right|$,\footnote{It should be noted that the local bias-minimising rescaling might be non-unique. Generally speaking, it is better to rescale the shape parameters than the location parameters: If $\Theta = \R^2$ and the first coordinate is the location parameter (mean), then this can be achieved by choosing $\cA=(1,\R)=\{(x_1,x_2) \in \R^2:x_1=1\}$.}
\item[(5)] Set the bootstrapped risk estimator to
\begin{equation}\label{eq:boot}
\hat \rho^{\textrm{b}}:=R(F_{a^{\textrm b} \circ \hat \theta}(X)) .
\end{equation}
\end{enumerate}
\end{tcolorbox}
}
\caption{The general algorithm for bootstrapping bias reduction in an i.i.d.~setting. Input quantities are $B$, the size of the bootstrapping sample, and $\cA\subset \R^d$ the chosen set of possible local bias minimizers. 
For some practical applications it can be useful to  adjust the algorithm in several aspects: for example to account for heavy tails or for particularly small data sets implying high variance in the parameter estimates.  We present such adjustments in Section \ref{sec:4}.} 
\label{f:boot.algorithm}
\end{figure}

To illustrate how Algorithm~1 works, let us calculate the value of $a^{\textrm b}$ for two reference VaR cases that were already considered before, i.e. for plug-in normal and plug-in GPD estimators. Inspired by the Gaussian case, \eqref{eq:gaussian.map}, we  apply rescaling only to the scale parameters. We fix the rolling window length to $n=50$, use i.i.d. samples, and consider VaR at level $5\%$. In the normal case, we compute univariate $a^{\textrm b}>0$ and get the adjusted plug-in estimator
\[
-\left(\hat\mu +(a^{\textrm b}\cdot \hat\sigma) \Phi^{-1}(0.05)\right),
\]
for various choices of true $\mu\in\bR$ and $\sigma>0$. In the GPD case we consider an exogenous threshold $u$ which we set to $u=0$ and consider univariate $a^{\textrm b}>0$ for the adjusted plug-in estimator
\[
\tfrac{(a^{\textrm b}\cdot \hat\beta)}{\hat\xi}\bigg(\alpha^{-\hat\xi}-1\bigg),
\]
for various exemplary choices of underlying true parameters $\xi\in\bR\setminus\{0\}$ and $\beta>0$. Note that while a linear adjustment $a^{\textrm b}$ is applied in the GPD case, it effectively takes into account values of parameter $\xi$ as the size of $a^{\textrm b}$ is computed locally for each $(\xi,\beta)$ and effectively depends on $\xi$.

\begin{figure}[tp!]
\begin{center}
\includegraphics[width=0.45\textwidth]{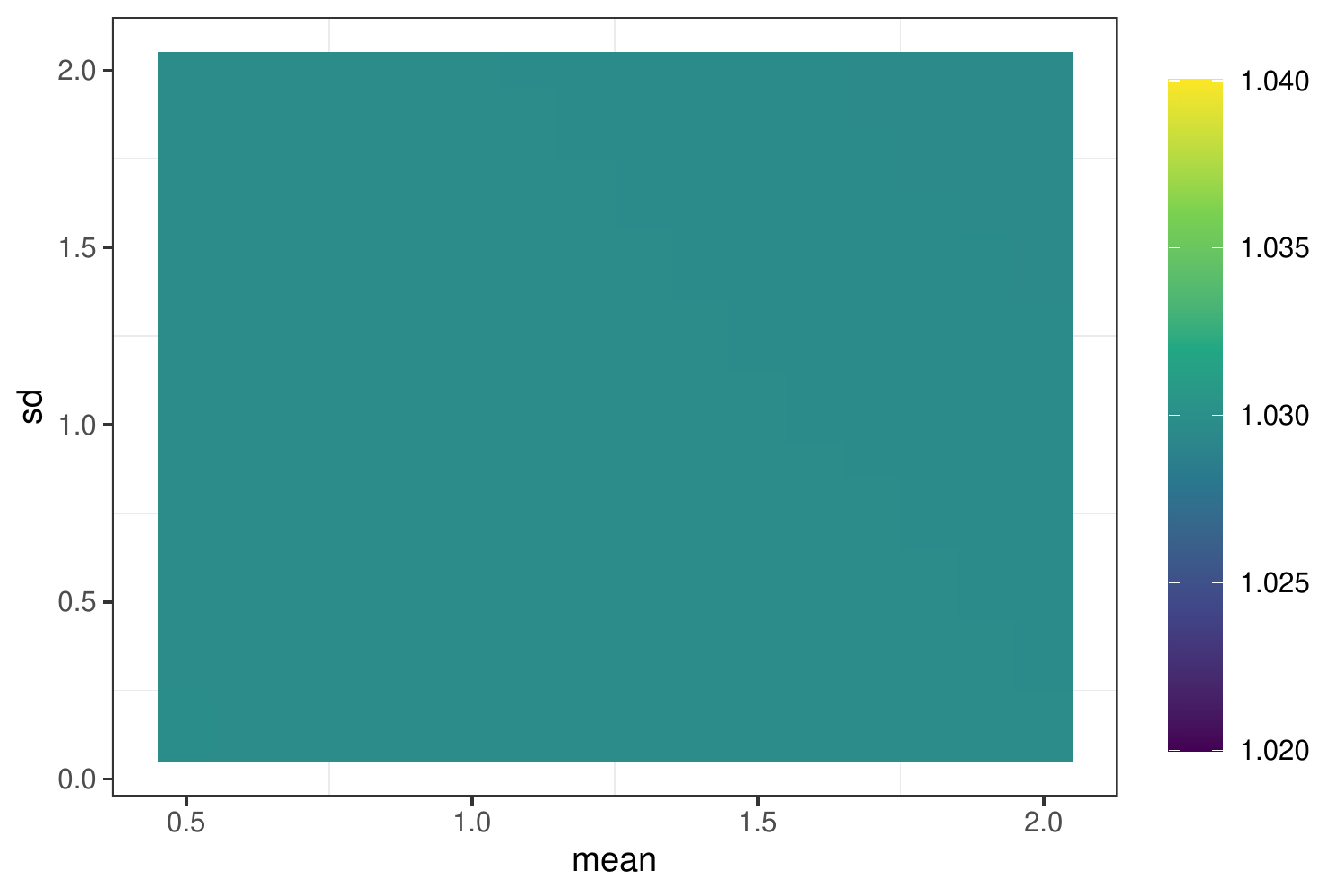}
\includegraphics[width=0.45\textwidth]{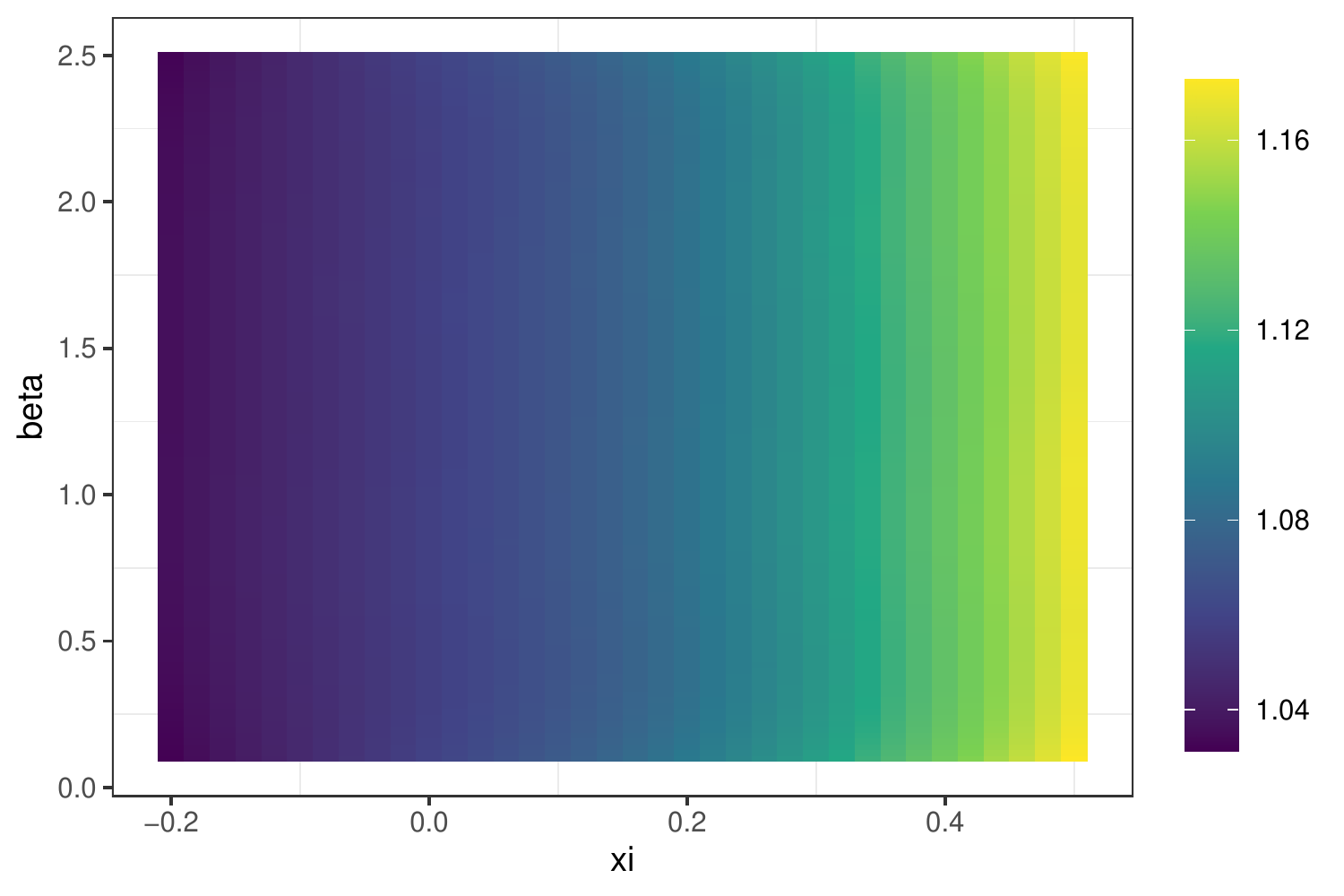} 
\end{center}
\caption{The heat plots present local bias adjustment  values for Normal (left) and GPD (right). We computed the multiplicative rescaler for scale parameters that should be applied to plug-in normal and GPD estimators to make them locally unbiased. The results are presented for sample size $n=50$ and VaR at level $p=5\%$.}
\label{F:local1}
\end{figure}

Also, in the GPD case we are using conditioned samples, i.e. 5\% denotes the conditioned VaR level. For example, assuming that initial sample size was 250 and 50 observations were negative, this setting would effectively correspond to unconditional threshold $5\%\cdot \tfrac{50}{250}=1\%$. In both cases, we set $B=50\,000$ for bootstrap sampling, and apply the algorithm detailed in Figure \ref{f:boot.algorithm}  on an exemplary representative parameter grid. The results are presented in Figure~\ref{F:local1}. 

Two main conclusions can be made at this stage. First, the local bias adjustment for the normal case does not depend on the underlying parameters and is close to the value presented in \eqref{eq:gaussian.map}, i.e. we have
\[
\sqrt{\tfrac{n+1}{n}}t_{n-1}^{-1}(p) / \Phi^{-1}(p)\approx 1.029.
\]
Second, in the GPD case, it could be observed that the bias value adjustment looks like an increasing function of $\xi$. This is in line  with intuition as the $\beta$ parameter is used for linear scaling which should not impact (relative) bias adjustment due to positive homogeneity of VaR.

\section{Performance of the local bias minimization}\label{sec:4}
In this section we present two examples of the local bias minimization procedures when the underlying returns follow a GPD distribution. For the first case we consider value-at-risk, while for the second we consider expected shortfall.


To asses the performance of the bias reduction, we simulate samples from the GPD families with different parameter settings and compare the performance of the plug-in estimators with its locally unbiased equivalent constructed via a variant of Algorithm 1. For completeness, in both cases we also show the results for {\it true risk} and {\it empirical} estimators.

The parameter sets we consider are 
\begin{align}
    \hspace{3cm}\theta_0 &= (-0.978,&&\hspace{-2.5cm}0.212,&&\hspace{-2.5cm}0.869), \notag\\
    \theta_1 &= (-2.2,&&\hspace{-2.5cm}0.388,&&\hspace{-2.5cm}0.545), \label{eq:parameters} \\  
    \theta_2 &= (-0.40028, &&\hspace{-2.5cm}1.19,&&\hspace{-2.5cm}0.774), \notag
\end{align}
while the $\alpha$ confidence thresholds are equal to $5\%$, 7.5\%, and $10\%$, respectively. In first two cases the learning period is set equal to $n=50$, while in the last case it is equal to $n=42$. 

The first setting corresponds to the Student $t$-distribution with 3 degrees of freedom and 20\% percentage threshold; the second setting corresponds to parameters calibrated to market data from the S\&P500 index -- see Table 5 in \cite{ManKel2006}; the third setting comes from Table 5 in \cite{Mos2004} and is related to an operational risk framework. Note that in the third set we have $\xi>1$ which implies that the first moment is infinite. Moreover, in the third case $n=42$ corresponds to the Corporate Finance data set from \cite{ManKel2006}, where $n=42$ peaks over the chosen threshold were observed. 

For simplicity, in all cases, we fix $u$, estimate $\beta$ and $\xi$, and rescale only $\beta$ when Algorithm 1 is used. 
The bootstrap sample size was chosen  to be $B=50\,000$, see Step (2) of Algorithm 1.

Already at this point it is visible that the peaks-over-threshold setting implies a very small number of observations: we have typically around 50 (or even only 42) data points at hand. Given that the underlying distribution has heavy tails (and in the third case additionally a non-existing first moment), a high variance of the parameter estimates and the presence of outliers can be expected. While in theory this could be remedied by increasing $n$, we propose to adjust the bootstrapping algorithm to account for this difficulties and present the adjusted procedure in the following. If the sample size $n$ is larger and the underlying distribution has light tails, one may safely proceed as in \cite{PitSch2016}, or rely directly on the bootstrapping algorithm presented in Figure \ref{S:boostrap.alg}.

To measure the performance of both VaR and ES estimators, we introduce a number of performance metrics. Recall that our inputs are risk estimators $(\hat\rho_t)$, realised P\&Ls $(x_t)$, and the corresponding secured positions $(y_t)=(x_t+\hat\rho_t)$.

\subsection{Performance metrics for value-at-risk}
For measuring the performance of estimators in the context of value at risk we consider the following five metrics: 
 first, the {\it exception rate}  $T$ defined in \eqref{eq:T.def}, i.e.
\[
T:=\frac 1 m \sum_{t=1}^{m}\1_{\{y_t<0\}}.
\]
Intuitively, here one would expect that $T$ should be close to the reference risk level chosen for the value-at-risk to be estimated.

Second, we consider the mean score statistic $S$ based on the quantile strictly consistent scoring function $s(r,x):=(\1_{\{r> x\}}-\alpha)(r-x)$, where $-r$ corresponds to the risk estimator, $x$ denotes the realised P\&L, and $\alpha$ is the underlying risk level. In our language, the secured position equals $y=x-r$ such that $\{r>x\}=\{x-r<0\}=\{y<0\}$ and we obtain the \emph{mean score}
\[
S:=-\frac{1}{m} \sum_{t=1}^{m}(\1_{\{y_t< 0\}}-\alpha)y_t.
\]
The mean score $S$ is a common elicitable  backtesting statistic used for the comparison of estimator performance, see~\cite{Gneiting2011} and \cite{Fissler2016}.

Third, we consider the rolling window traffic-light {\it non-green zone} backtest statistic NGZ as already mentioned. We fix the length of an additional rolling window to $N=50$. 
The  number of overshoots in the interval $[s,s+N-1]$ equals $T_1(s):=\sum_{t=s}^{s+N-1}\1_{\{y_t<0\}}$ and
the average {\it Non-Green Zone} (NGZ) classifier is defined as 
\[
\textrm{NGZ}:=\frac{1}{m-N}\sum_{s=1}^{m-N}\1_{\{T_1(s)\geq z_{\alpha}\}},
\]
where the exception number $z_\alpha$ is the  95\% confidence threshold. Namely, under the correct model, the sequence $(\1_{\{y_t<0\}})$ is i.i.d.~Bernoulli distributed  with probability of success equal to $\alpha$. Consequently,  $T_1(t_0)$ should be linked to a Bernoulli trial of length $n$ with success probability $\alpha$. For example, for $\alpha=0.05$ and $n=50$, we get 
$$
    F_{B(50,0.05)}(4)\approx 0.89 \quad \text{ and }\quad F_{B(50,0.05)}(5)\approx 0.96,
$$
 which would lead to 95\% confidence threshold equal to $z_\alpha=5$. Hence, in a perfect setting, NGZ should be close to $1-F_{B(50,0.05)}(z_\alpha-1)\approx 11\%$ for the first dataset. Using similar reasoning, for the second and the third datasets settings, we get NGZ values equal to approximately 8\% and 5\%, respectively. 

Fourth, we introduce \emph{Diebold-Mariano} comparative ($t$-)test statistic with unbiased estimator being the reference one.  The test statistic is given by
\[
 \textrm{DM}:= \sqrt{n}\,\cdot \frac{\hat\mu (d) }{\hat\sigma(d)},
\]
where $d=(d_t)_{t=1}^{m}$ is the time series of comparative errors between two estimators,  $\hat \mu(d)$ is the sample mean comparative error, and  $\hat\sigma(d) $ is the sample standard deviation of the comparative error, respectively; see~\cite{osband1985information} for details.  In the VaR case, the $t$-th day comparative  error is given by 
$$
    d_t:=s(-\hat \var^{z_1}_t,x_t)-s(-\hat \var^{z_2}_t,x_{t}), 
$$
where $z_1$ and $z_2$ denote the considered estimators (such as plug-in estimator, empirical estimator, or true risk) and $s(r,x)=(\1_{\{r> x\}}-\alpha)(r-x)$ is the consistent scoring function already introduced above.

Finally, we consider the mean and the standard deviation of the estimated regulatory capital. When all other criteria are satisfied, these additional criteria allows to identify the estimator which requires minimal regulatory capital and/or has the smallest statistical bias. We  define the {\it mean of risk estimator} (MR) and \emph{standard deviation of risk estimator} (SD) by 
\begin{align}\label{def:MRV}
\textstyle \textrm{MR}:= \frac{1}{m}\sum_{t=1}^{m} \hat\rho_{t}\quad\textrm{and}\quad  \textrm{SD}:= \sqrt{\frac{1}{m-1}\sum_{t=1}^{m} (\hat\rho_{t}-\textrm{MR})^2}.
\end{align}
Note that smaller values of MR indicate that (on average) smaller capital reserves are required. Ideally, this metric should be close to the true risk. Since parameters need to be estimated we expect that this model uncertainty is reflected by an increase in MR. 

\subsection{Performance metric for expected shortfall}

For simplicity, we decided to focus only on the backtesting performance; other conclusions should be similar to VaR. In this regard, we consider  the averaged cumulative exception rate defined in \eqref{eq:G.def}, i.e.
\[
 G:=\frac{1}{m}\sum_{i=1}^{m}1_{\{y_{(1)}+\ldots+y_{(i)}<0\}}.
\]
For completeness, we also include results of MR and SD in the ES case.



\subsection{Bias reduction for value-at-risk in the GPD framework}\label{ex:31}
For the following data experiment, we consider the three different sets of GPD parameters $(u,\xi,\beta)$ and estimate VaR  using three different (conditional) reference $\alpha$ levels.  We follow a backtesting rolling window approach with total number of simulations equal to $N=n+m=100\,000$. 
We consider five different VaR estimators: the empirical estimator (emp), the plug-in GPD estimator (plug-in), the true risk (true), and two bootstrap-based estimators which will be described below.

First, to obtain a suitable reference benchmark for the bootstrapping procedure, we adjust the GPD plug-in estimator $\hat\var^{\textrm{plug-in}}$ by a fixed multiplier $a$ that is computed based on \emph{true values} of $\xi$ and $\beta$.  
In this regard, set
\begin{equation}\label{eq:unb1}
\hat \var^{\textrm{b,true}}_t:= -u+a^{\textrm{b,true}}\cdot (\hat\var^{\textrm{plug-in}}_t+u),
\end{equation}
where $a^{\textrm{b,true}}$ is computed using true values of $\xi$ and $\beta$ in the first step of Algorithm 1; note that adjusting $\hat\beta$ by $a$ results in \eqref{eq:unb1}.

Second, we consider an estimator that is computed using the bootstrapping algorithm detailed in Figure \ref{f:boot.algorithm} applied to each sample separately. We set
\begin{equation}\label{eq:unb3}
\hat \var^{\textrm{b}}_t:= -u+ a^{\textrm b}_{t}\cdot (\hat\var^{\textrm{plug-in}}_t+u),
\end{equation}
here $a^{\textrm b}_{t}$ is the local bias correction for the estimates $\hat\xi_t$ and $\hat\beta_t$ on day $t$.

 \begin{table}[t!]
\caption{The table presents the estimation of value-at-risk under GPD distributions.
It reports the true value-at-risk $\hat \var_t^{\textrm{true}}$ as benchmark and the estimators $\hat \var_t^{z}$ with $z\in\{\textrm{emp}, \textrm{plug-in}, \textrm{b-true},\textrm{b}\}$ together with the exception rate $T$, the mean score $S$, the non-green zone (NGZ), the Diebold-Mariano test statistic with $\hat \var_t^{\textrm{b-true}}$  as reference  (with double-sided $p$-values), and the mean risk value MR (with standard deviation SD in brackets). The bootstrapped estimator $\hat \var_t^{\textrm{b}}$ outperforms all other estimators (except for the benchmark of course). 
}
\centering

\begin{tabular}{llccccrrr}\toprule
Parameters      & VaR& $\alpha$ & $n$                                    & $T$   & $S$  & NGZ  & DM ($p$-value)   & MR (SD)   \\ \midrule
GPD             & emp       &\multirow{5}{*}{0.050}&\multirow{5}{*}{50}& 0.066 & 0.2475 &0.20 & -11.8 (0.000)& 4.47 (0.92)\\
$u=-0.978$      & plug-in   &&                                         & 0.060 & 0.2426 &0.15 & -0.4  (0.689)& 4.57 (0.78)\\
$\xi=0.212$     & b-true    &&                                         & 0.052 & 0.2425 & {0.08} & ---    & 4.83 (0.83)\\
$\beta=0.869$   & b &&                                             & 0.052 & 0.2424& {0.08} & {2.2} (0.028) & 4.83 (0.80)\\
                & true      &&                                         & \textbf{0.051} & \textbf{0.2331}& \textbf{0.11} & \textbf{18.7}  (0.000) & 4.61 (0.00)\\ 
 \midrule

GPD           &emp&\multirow{5}{*}{0.075}&\multirow{5}{*}{50}  & 0.091 & 0.3140 & 0.13 & -10.7 (0.000)& 4.59 (0.71)\\
$u=-2.200$    &plug-in  &&                                & 0.087 & 0.3101 &0.12 & -0.5   (0.617) & {4.58} (0.57)\\
$\xi=0.388$   &b-true &&                                   & 0.078 & 0.3100 & {0.07}& ---   & 4.76 (0.61)\\
$\beta=0.545$ &b  &&                           &{0.077} &0.3097 & {0.07} & {4.4}    (0.000)  & 4.76 (0.58)\\
              &true  &&                               & \textbf{0.077} & \textbf{0.3017} & \textbf{0.08} & \textbf{18.2}   (0.000)  & 4.63 (0.00)\\ 
 \midrule

GPD          &emp &\multirow{5}{*}{0.100}&\multirow{5}{*}{42}& 0.119 &7.2509 & 0.08 & -13.3  (0.000)  &10.57 (7.16) \\
$u=-0.400$   &plug-in  &&                                 & 0.124 &7.2003 & 0.12 & 0.0   (1.000)  & {9.08} (4.51) \\
$\xi=1.190$  &b-true&&                                 &0.111 &7.2003 & 0.07 & ---  &10.38 (5.19) \\
$\beta=0.774$&b &&                             &0.111 &7.1907 & 0.07 & \textbf{13.2}    (0.000)  &10.30 (4.60) \\
             &true &&                                 & \textbf{0.101} & \textbf{7.1175} & \textbf{0.05} & 12.0  (0.000)  & 9.82 (0.00) \\ 
 \bottomrule
\end{tabular}
\label{T:table1}
\end{table}

As already mentioned, in the case with small sample size ($n=50$ or $n=42$) and the presence of heavy tails it is necessary to adjust Algorithm 1 for this. In this regard, we proceed as follows:

\begin{enumerate} 
    
    \item To provide a challenging benchmark $\hat \var^{\textrm{b,true}}_t$ we realized that in the small sample context it turns out to be useful to allow a small positive bias of the estimator. In this regard, we modified objective function in step (4). Namely, in all cases we allowed a bias equal to 10\% of the estimator standard error. For the first two datasets, this corresponds to approximately 1\% of the true risk value, while for the third dataset this corresponds to 10\% of the true risk value. Note that in the third case we have $\xi>1$ which increases the standard error significantly. 
    
    \item For the bootstrapping estimator $\hat \var^{\textrm{b}}_t$ we only took care of outliers: namely, we set $ a^{\textrm b}_t\equiv 1$  (i.e. do not apply any adjustment) if the estimated value of $\hat\var^{\textrm{plug-in}}_t$ seemed already very high. We chose the case where the estimated plug-in risk was above the  10\% upper quantile of the aggregated estimated capital sample $(\hat\var^{\textrm{plug-in}}_t)$. This corresponds to situations, where the size of non-adjusted (above the threshold) estimated risk is already higher than the (above the threshold) true risk by approximately 28\% for the first two datasets, and higher than 48\% for the third dataset.
\end{enumerate}


 The results of the performance analysis are presented in Table~\ref{T:table1} and we can make the following observations.
 
\begin{enumerate}
	\item The bootstrapped estimator $\hat \var^{\textrm{b}}$ shows its unbiasedness by reaching an exception rate $T$ that is closest to the true risk exception rate. Its values of the mean score $S$ and the non-green zone NGZ are also closest to the true risk. Moreover, it outperforms other estimators in the Diebold-Mariano test.
	\item The modification (2) above, applied to the bootstrapping estimator $\hat \var^{\textrm{b}}$ led to the outperformance over the  estimator $\hat\var^{\textrm{b,true}}$ shown in the Diebold-Mariano test statistics (DM). This shows that in the heavy-tail- and small-sample-environment one needs to pay special attention to the bootstrapping bias correction. Intuitively, one needs to robustify the estimating procedure and in particular one should  avoid adjusting estimates which  already considerably overestimate the underlying risk. 
	
	\item The mean risk of the bootstrapped estimators is typically higher compared to the other estimators showing that biased estimators often underestimate risk. 
	\item In the third, extreme case we observe that the empirical estimator performs quite well in terms of the no-green-zone (but not in terms of the other measures). This results from a high capital requirement shown in the mean risk statistic MR. 
\end{enumerate} 
 

Summarizing, the above results show that the suggested improved bootstrapping procedure performs very well even in the difficult context considered here with a very small sample size and in the presence of heavy tails. It also complements the asymptotic analysis presented in Section~\ref{ex:VAR.bt} showing that the plug-in procedure indeed underestimates  risk in the setting considered here.


\subsection{Bias reduction for expected shortfall in the GPD framework}\label{ex:32}

\begin{table}[tp!]
\caption{
The table presents numerical results the estimation of expected shortfall in two cases of a GPD distribution.
It reports the empirical (emp), the plug-in, the estimator knowing $\xi$ and $\beta$ (true), the bootstrapping estimator using $\xi$ and $\beta$ for computing $a$ (true2), and the two locally unbiased estimators: the bootstrapping estimator (boot) and the splitting estimator (splitting). 
The reported values are the cumulative aggregated exception rate statistic $G$, and the mean risk value MR (with standard deviation in brackets).}

\centering

\begin{tabular}{l|l|c|c|c|r}\toprule
Parameters & ES& $\alpha$ & $n$                         & $G$     &  MR (SD)   \\ \midrule
GPD           &emp      &\multirow{5}{*}{0.050}&\multirow{5}{*}{50}& 0.100 & 6.14 (1.83) \\
$u=-0.978$    &plug-in  &&                                & 0.078 & 6.73 (2.01) \\
$\xi=0.212$   &b-true&&                                    & 0.050 & 7.88 (2.41) \\
$\beta=0.869$ &b&&                                    & 0.057 & 7.66 (2.22) \\
              &true &&                                    & \textbf{0.051} & 6.70 (0.00) \\ 
 \midrule

GPD             & emp&\multirow{5}{*}{0.075}&\multirow{5}{*}{50}  & 0.136 & 6.80 (2.40) \\
$u=-2.200$      &plug-in  &&                                  & 0.121 & 7.08 (2.44) \\
$\xi=0.388$     &b-true &&                                     & 0.079 & 8.35 (3.08) \\
$\beta=0.545$   &b &&                                     & 0.092 & 8.03 (2.65) \\
                &true &&                                      & \textbf{0.077} & 7.07 (0.00) \\ 
 \bottomrule
\end{tabular}
\label{T:table2}
\end{table}

In this section we consider expected shortfall instead of value-at-risk. Expected shortfall is much more sensitive to heavy tails, such that a priori we can expect an even clearer picture. 

To this end, we consider the first two parameter sets from Equation \eqref{eq:parameters}. Note that in the third dataset we have  $\xi>1$ which implies an exploding first moment, such that expected shortfall is no longer finite. The results are presented in Table~\ref{T:table2} and the following observations can be made:

\begin{enumerate}
	\item The bias adjustment for expected shortfall is large in comparison to the adjustment in the value-at-risk case, computed for the same confidence threshold $\alpha\in (0,1)$. This is visible through the increased mean risk MR. 
	\item The performance (measured in terms of $G$) for empirical and plug-in estimators is significantly  worse in comparison to the bootstrapped estimator, since the former  tend to underestimate the risk. This is in line with the results presented for the VaR case.
\end{enumerate}

\begin{figure}[tp!]
\begin{center}
\includegraphics[width=0.45\textwidth]{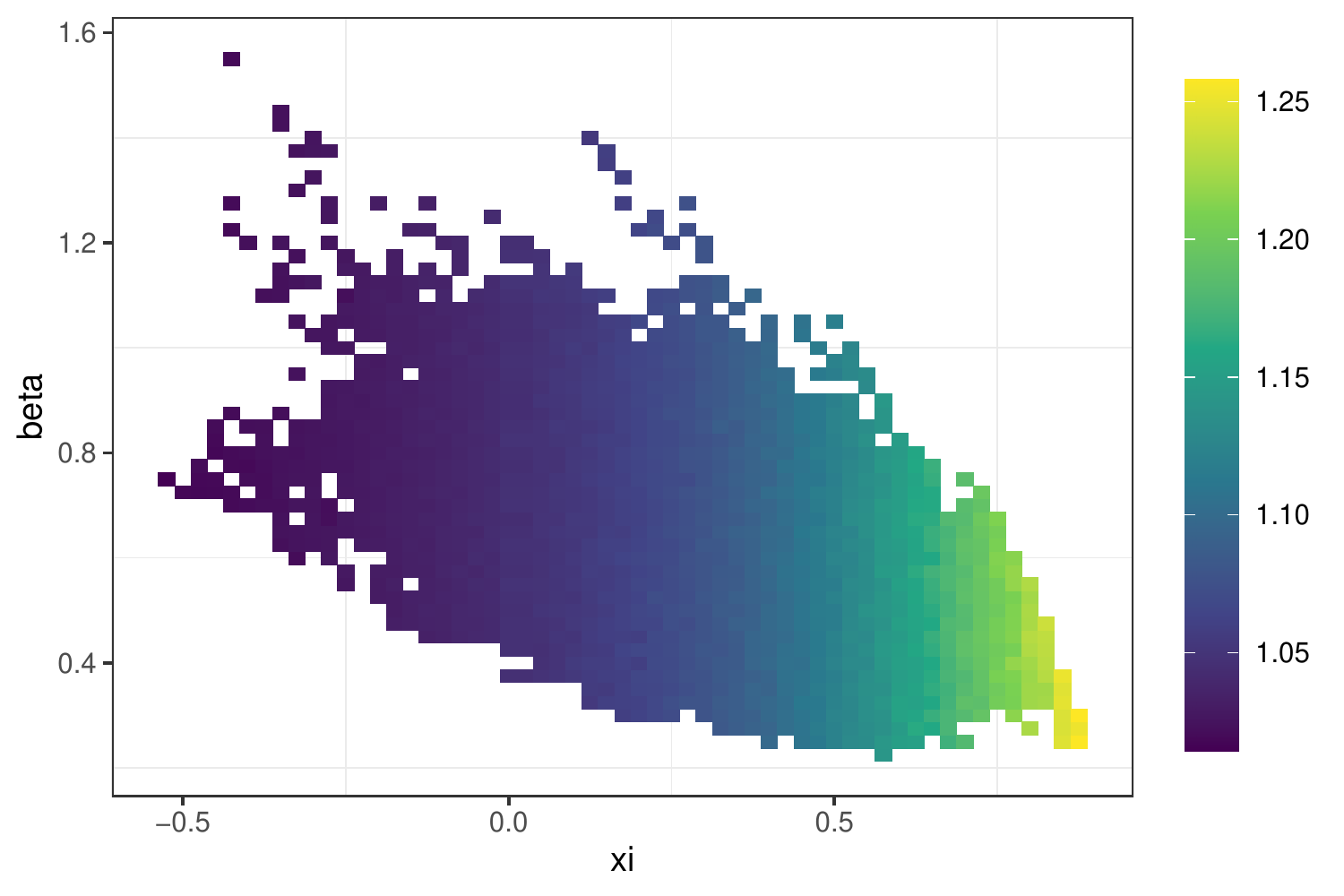}
\includegraphics[width=0.45\textwidth]{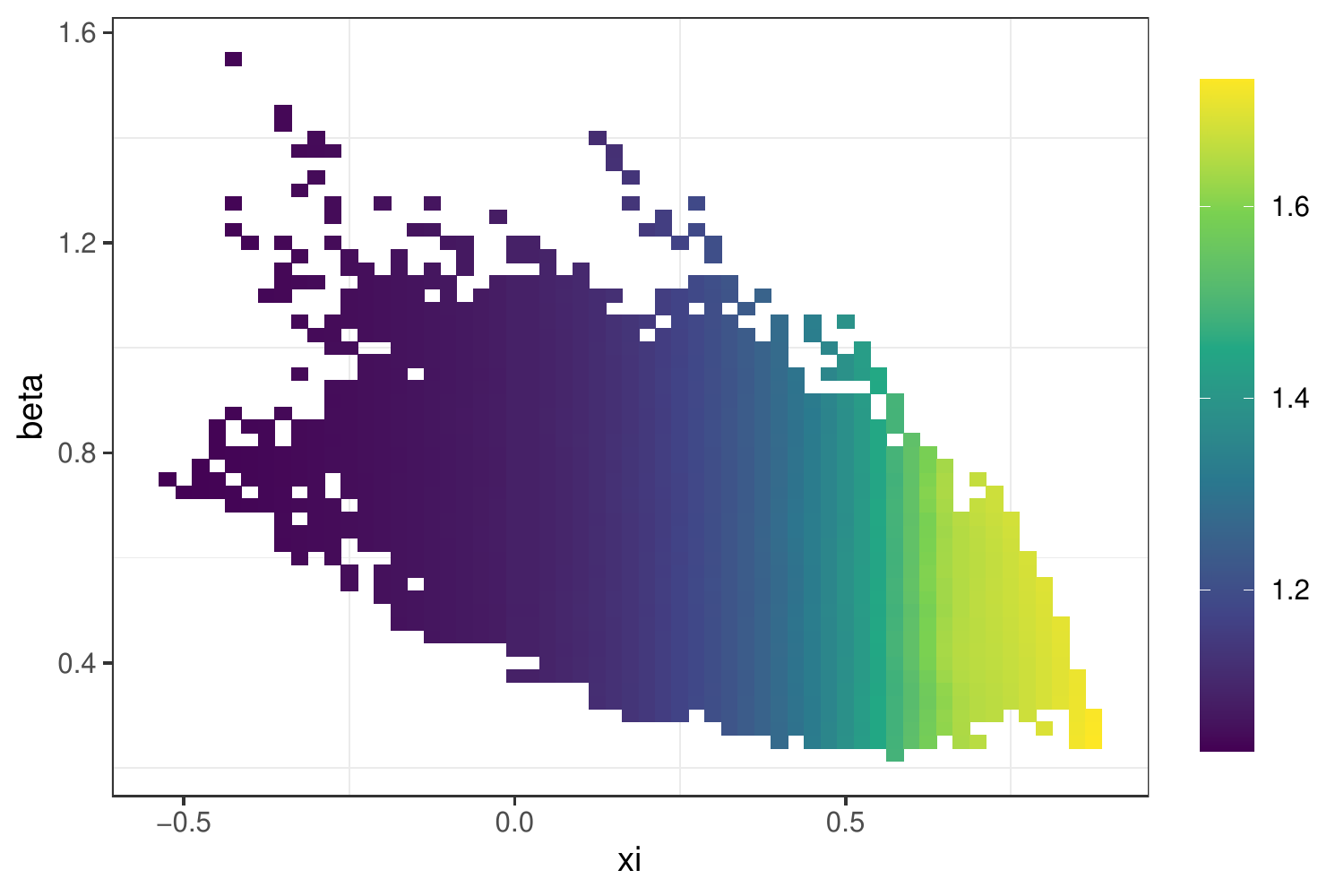} 
\end{center}

\vspace{-5mm}

\caption{The heatplots present local bias adjustment for GPD plug-in estimator for VaR (left) and ES (right), both at level 7.5\%. The underlying sample size in both cases is equal to $n=50$. The multiplicative rescalers are computed for all parameter pairs that were estimated for 100\, 000 simulated data based on Dataset 2 specification.}
\label{F:local2}
\end{figure}

Additionally, we illustrate the effect from observation (1) with a plot showing value-at-risk and expected shortfall  bias adjustments computed for all estimated parameter values for the second dataset, see Figure~\ref{F:local2}. Note that the these results are in perfect agreement with results in Figure~\ref{F:local1}, i.e. $\xi$ is the main bias determination driver.

\section{Conclusion}\label{S:conclusions}


Our experiments show that plug-in estimators of risk capital typically suffer from an underestimation of risk. We could show that this effect is more pronounced when heavy tails are present, the sample size is small or heteroscedasticity drives the underlying process. This underestimation is measurable by backtests, and we analyzed a number of backtests related to value-at-risk, expected shortfall or risk measures based on expectiles. Moreover, we suggest a new bias reduction technique based on a bootstrap procedure which increases the efficiency of plug-in estimators.  Our findings highlight  that predictive inference in reference to the estimation of risk and risk bias is an interesting topic that  requires further studies, especially in  non-i.i.d.\ settings.

\section*{Acknowledgements}
The first author acknowledges support from the National Science Centre, Poland, via project 2016/23/B/ST1/00479.
The second author acknowledges support from the Deutsche Forschungsgemeinschaft under the grant SCHM 2160/13-1.

 {\small
 \bibliographystyle{agsm}
 \bibliography{bibliography}
 }

 \end{document}